\documentstyle[12pt,aasms4]{article}	

\def\Section#1{\S #1}

\slugcomment{Submitted to Astronomical Journal.}

\lefthead{Scoville et al.}
\righthead{M51}

\newcommand{\etal}{\mbox{\it et~al.\ }}

\newcommand{\msun}{\mbox{$M_{\odot}$}}

\newcommand{\lsun}{\mbox{$L_{\odot}$}}
\newcommand{\kms}{\mbox{km s$^{-1}$}}

\def\deg      {{\ifmmode^\circ\else$^\circ$\fi} } 

\def\h2     {H$_2$}

\begin{document}

\title{High Mass, OB Star Formation in M51 :\\ 
HST H$\alpha$ and P$\alpha$ Imaging}

\author{N. Z. Scoville}
\affil{California Institute of Technology, Pasadena, CA 91125}

\author{M. Polletta}
\affil{California Institute of Technology, Pasadena, CA 91125
\and Observatory of Geneva, Sauverny, Switzerland}

\author{S. Ewald, S. R. Stolovy }
\affil{California Institute of Technology, Pasadena, CA 91125}

\author{R. Thompson and M. Rieke}
\affil{Steward Observatory, University of Arizona, Tucson, AZ 85721}

\begin{abstract}
We have 
obtained H$\alpha$ and P$\alpha$ emission line images covering the 
central 3 --- 4\arcmin ~ of M51 using the WFPC2 and NICMOS cameras on HST to study the high-mass stellar population. The 0.1 --- 0.2\arcsec\ pixels provide 4.6 --- 9 pc resolution in M51 and  the H$\alpha$/P$\alpha$ line ratios are used to obtain extinction estimates. A sample of 1373 H$\alpha$ emission regions is catalogued using an automated and uniform measurement algorithm. Their sizes are typically 10 --- 100 pc. The luminosity 
function for the H$\alpha$ emission regions is obtained over the range  L$_{H\alpha}$ =
10$^{36}$ to 
$2\times 10^{39}$ erg s$^{-1}$. The luminosity function is fit well
by a power law with $dN/dlnL \propto L^{-1.01}$).  The power law is significantly 
truncated and no regions 
were found with {it observed} L$_{H\alpha}$ above 2$\times 10^{39}$ erg s$^{-1}$ (uncorrected for extinction). (The maximum seen in ground-based studies is approximately a factor of 5 higher, very likely due to blending of multiple regions.) The extinctions 
derived here increase the maximum {\it intrinsic} luminosity to above 10$^{40}$ erg s$^{-1}$).
The logarithmically binned luminosity function is also somewhat steeper ($\alpha = -1.01$) than that found ground-based imaging ($\alpha = -0.5 \to -0.8$)  ---  
probably also a result of our resolving regions which were blended in the ground-based images. The 2-point correlation function for the HII regions
exhibits strong clustering on scales $\leq 2$\arcsec\ or 96 pc.

To analyze the 
variations of HII region properties {\it vis-a-vis} the galactic structure,
the spiral arm areas were defined independently
from mm-CO and optical continuum imaging. Although the arms
constitute only 25\% of the disk surface area,  the arms contain 45\% of the catalogued HII regions.
The luminosity function is somewhat flatter in spiral arm regions than in the interarm areas (-0.72 $\rightarrow$ -0.95); however, this is very likely the result of 
increased blending of individual HII regions in the arms which have higher surface density. No significant difference is seen in the  sizes and electron densities of the HII regions in spiral arm and interarm regions. For 209 regions which had $\ge$ 5$\sigma$ detections in both P$\alpha$
and H$\alpha$, the observed line ratios indicate visual extinctions in the range  A$_V$ = 0 to 6 mag. The mean extinction was A$_{V}$ = 3.1 mag
(weighting each region equally), 2.4 mag (weighting each by the 
observed H$\alpha$ luminosity) and 3.0 mag (weighting by the extinction-corrected luminosity). On average, the observed H$\alpha$ luminosities 
should be increased by a factor of $\sim 10$, implying  comparable increases in 
global OB star cluster luminosities and star formation rates.  The full range of extinction-corrected H$\alpha$ luminosities
is between $10^{37}$ --- 2$\times$10$^{40}$ erg s$^{-1}$.

The most luminous regions have sizes $\ge 100$ pc so it is very likely 
they are blends of multiple regions. This is clear based on their sizes
which are much larger than the maximum diameter ($\leq 50$ pc) to which an HII region might conceivably 
expand within the $\sim 3\times 10^6$ yr lifetime of the OB stars. It is 
also consistent with observed correlation (L $\propto D^{2}$) found between the 
measured luminosities and sizes of the HII regions. We therefore 
generated a subsample of 1101 regions with sizes $\leq 50$ pc which 
constitutes those region which might conceivably be ionized by a single cluster.
Their extinction-corrected luminosities range between $2\times 10^{37}$ and $10^{39}$ erg s$^{-1}$, or between 2/3 of M42 (the Orion Nebula) and W49 (the most luminous Galactic radio HII region).
The upper limit for 
individual clusters is therefore conservatively $\leq 10^{39}$ erg s$^{-1}$, implying Q$_{LyC}$~$_{up} \simeq 7\times10^{50}$
s$^{-1}$ (with no corrections for dust absorption of the Lyman continuum 
or UV which escapes to the diffuse medium). This corresponds to cluster masses $\leq $5000 \msun (between 1 and 120 \msun). 

 The total star formation 
rate in M51 is estimated from the extinction-corrected H$\alpha$ 
luminosities to be $\sim$ 4.2 \msun ~yr$^{-1}$ (assuming a Salpeter IMF between 
1 and 120 \msun) and the cycling time from the neutral ISM into
these stars is $1.2\times 10^9$ yr.  

We develop a simple model for the UV output from OB star clusters
as a function of the cluster mass and age in order to interpret 
constraints provided by the observed luminosity functions. The power-law index 
at the high luminosity end of the luminosity function ($\alpha = -1.01$) implies
N(M$_{cl}$)/d~M$_{cl} \propto M_{cl}^{-2.01}$. The high
mass clusters ($\sim$ 1000 \msun) have a mass such that the IMF is well sampled up to $\sim$ 120 \msun, 
but this cluster mass is $\leq$1\% of that available in a typical GMC. We 
suggest that OB star formation in a cloud core region is 
terminated at the point that radiation pressure on the 
surrounding dust exceeds the self-gravity of the core star cluster and that this is what limits the maximum 
mass of standard OB star clusters. 
This occurs at a stellar luminosity-to-mass ratio $\sim$ 500 -- 1000 \lsun/\msun~ 
which happens for clusters $\geq 750$ \msun. We have modelled the core collapse hydrodynamically and find that a second wave of star 
formation may propagate outwards in a radiatively compressed shell 
surrounding the core star cluster --- this triggered, secondary star formation
may be the mechanism for formation of super star cluster (SSC) seen in 
starburst galaxies.

\end{abstract}


\section{Preamble}

Were it not for the small number of youthful, luminous stars and 
their ongoing genesis, much of the beauty, vigor and evolution that is our fascination in the universe would be lost to the distant past. Energizing and enriching the disks of galaxies are the most massive stars of each 
generation. In youth, they illuminate the bright nebulae which so elegantly 
outline the spiral arms of distant galaxies; in death, their cataclysmic supernovae replenish the interstellar environment with gases, enriched 
in heavy elements. From these ashes the future generations of stars
will arise. The springs of rejuvenation are giant molecular clouds
encompassing millions of solar masses of cold gas. Inside these ponderous 
cocoons, the metamorphosis of stars takes place in dusty darkness. 

\section{Introduction}

High mass OB stars play a critical role in the energetics and dynamics
of the ISM and in the highest luminosity phases of galactic structure
and evolution, specifically the luminous spiral arm and starburst
activity. Nevertheless, the mechanisms for formation
of OB associations remain poorly understood and indeed, it is 
uncertain whether high and low mass stars are formed by the same 
or different processes (e.g. \cite{lar86}). HII regions have long been a primary probe of 
high-mass star formation and the properties of OB star
clusters (\cite{hod87}, \cite{ken89}, \cite{ran92}, \cite{thi01}). The hydrogen recombination line fluxes (eg. H$\alpha$) or 
radio free-free continuum are proportional to the volume-integrated emission measure of the
HII
regions. The latter is proportional to the 
total Lyman continuum emission rate of the associated high-mass stars
under the assumption that all ionizing photons are locally absorbed.
Thus the H$\alpha$ luminosity of an emission region is indicative 
of the Lyman continuum emission and hence 
the mass of high-mass stars (correcting for extinction and assuming an IMF). The luminosity function
of the HII regions can then be used to study the distribution 
of masses and birth rates of OB associations. M51 is a Rosetta stone  for studies of OB star formation  ---  on account of its proximity  ---  9.6 Mpc (Sandage and Tammann 1975); its grand design spiral pattern; its orientation  ---  i = 20\deg (\cite{tul74}) and its abundant, dense ISM (\cite{sco83}).  

M51 has been the focus of numerous ground-based H$\alpha$ studies (\cite{ken89}; \cite{ran90};
\cite{van88}; \cite{ran92}; \cite{roz96}; \cite{pet96}; and \cite{thi00}),
and these studies have contributed much of what is currently known
regarding OB associations in other spiral galaxies. The distribution of HII region H$\alpha$ luminosities was approximately fit
by a truncated power-law with (N(L)~d~ln(L) $\propto$ L$^{-0.55 \to -0.75}$)  ---  on the high luminosity end. 
In several previous H$\alpha$ studies (eg. Rand 1992 and Thilker \etal 2000), the luminosity
function appears steeper in the interarm regions than in the arms (exponent $-$0.93 $\to$ $-$0.96 versus $-$0.48 $\to$ $-$0.72). The bright end (L$_{H\alpha} \geq 10^{38.8}$ erg s$^{-1}$) of the HII region luminosity function contributes the 
bulk ($\geq 50$\%) of the discrete HII region luminosity (\cite{ran92}), but approximately 55\% of the total 
H$\alpha$ luminosity originates from diffuse ionized gas (DIG), i.e. not in discrete 
regions (\cite{ran92}; \cite{gre98}). Based on ground-based imaging, it remains uncertain whether the 
DIG is blended low luminosity regions or truly diffuse gas.

We have recently completed a comprehensive study 
of M51 comprised of HST (WFPC2 and NICMOS) imaging of the optical and near infrared continuum (\cite{pol01}) and the H$\alpha$ and P$\alpha$ emission lines 
(this paper). Related mm-CO interferometry has been presented in Aalto \etal (1999). The former probes the 
stellar disk, the dust, and the OB star formation while the  
latter probes the dense, molecular ISM which is the birthsite of OB star clusters. 

The 0.1 --- 0.2\arcsec\ resolution available with HST imaging 
corresponds to 4.6 --- 9.3 pc. These sizes correspond to those of 
individual resolved Galactic HII regions (eg. the Orion Nebula). 
For comparison, ground-based
H$\alpha$ imaging at resolutions $\geq$ 1-2 \arcsec\ corresponds to at least 50  --- 100 pc, the size of a large Galactic giant molecular cloud (GMC).
The latter resolution will clearly blend multiple sites of OB star
formation which occur within a single GMC (for example M42 and NGC 2024 
in the Orion GMC). At the same time, the ground-based resolution 
element will also contain enormous volumes of intervening neutral gas.
The high angular resolution of the HST imaging is thus critical for the 
study of extragalactic HII region properties.

To date there have been surprisingly few studies of extragalactic HII 
regions using HST. A recent study of M101 by Pleuss \etal (2000) clearly demonstrates
the advantages of HST. Specifically, the break in the luminosity function (LF) slope at Log(L$_{H\alpha}$) = 10$^{38.6}$ seen in some ground-based studies (\cite{bec00} -- attributed to the transition
from ionization-bounded to density-bounded regions) was not apparent. And multiple HII regions were resolved into individual regions
of lower luminosity, resulting in a very different distribution of HII region sizes and enabling an analysis of HII region clustering (\cite{ple00}).

\subsection {Our Study}

  This paper which presents HII 
emission line imaging addresses the
following specific issues :

1) the global luminosity function of the HII regions and their associated OB star clusters;

2) comparison of 
the form of the luminosity function with theoretical expectations;

3) variations in the luminosity function from arm to interarm regions
and between the nuclear region and the galactic disk;

4) the HII region sizes and densities; and

5) analysis of the reddening and extinction of the HII regions based on
the observed ratios of H$\alpha$/P$\alpha$ lines.

To address these issues in a meaningful way and fully understand
the observational limitations, we develop automated routines for
the definition of the HII region boundaries and a model to simulate
the blending of multiple regions (\Section{5.1}). The 
temporal evolution of the Lyman continuum emission from an OB star cluster
is also modelled and constraints on the cluster mass distribution 
are derived from the observed luminosity function of HII regions. Lastly, we analyze the physical processes important in 
determining the masses and sizes of OB star clusters forming within a molecular cloud
core.

In the following sections, we present the observations and images for H$\alpha$ and P$\alpha$ (\Section{3-4}
) and discuss the definition and measurement of the  HII regions (\Section{5}
). The clustering of HII regions, in particular a 2-point correlation 
function is derived in \Section{6}. The observed H$\alpha$ luminosity funtion and Lyman continuum emission rates are presented in \Section{7 \& 8}. The size and density distributions for arm and interarm regions are discussed in \Section{9}.
The H$\alpha$/P$\alpha$ ratios are used to estimate extinctions of 209 HII regions in \Section{10} and hence derive extinction-corrected luminosities.  
In \Section{11} we present a sample of HII regions with size $\leq$ 50 pc and then compare these with Galactic HII regions in \Section{12}. 
In \Section{13} we present a simple model invoking instantaneous 
OB star formation with a standard IMF to provide a context for interpretation
of the observational data. We also develop a model for the formation 
of OB star clusters in which the mass of the core cluster is limited 
by radiation pressure once the cluster has accumulated $\sim$ 1000 \msun.
In \Section{14}, we use the total fluxes in H$\alpha$ to estimate 
the overall Lyman continuum production and star formation rate in M51.
 
\section{Observations}

HST imaging of M51 using both WFPC2 and NICMOS was obtained in several observing 
programs as discussed in Polletta \etal (2001) and summarized in Table 1. 

\subsection{WFPC2}

The WFPC2 continuum and
H$\alpha$ images were obtained in 1995 January (\cite{for96})
and in 1999 July by Scoville \& Ewald (filling in the areas not covered in the 
Ford \etal archive data).  The raw images were flat-fielded using the automatic standard pipeline at the STScI and cosmic rays were removed using our own procedure. The complete calibration procedures are described in Polletta \etal (2001).For subtraction of continuum from the F656N image, which includes H$\alpha$ and continuum, the Y-band (F547M) image was used in the earlier epoch and I-band (F814W) for the later epoch.  The broad band continuum 
image was first scaled by a constant such that the signal strengths on a sample of stars balanced those in the F656N image. 

To test that the use of different continuum filters for the two fields did not introduce 
photometric differences,
we compared the measured fluxes for regions in the overlap area of the separate images. In the 
H$\alpha$+continuum image the average difference was 1.6\% . The H$\alpha$ flux (after continuum subtration) was different by $\sim$2\% for faint regions and 6\% for the brightest regions.

\subsection{NICMOS}

NICMOS (187N and 190N) 
images were obtained as part of the NICMOS GTO program with a 
mosaic of 9 NIC3 fields covering most of the area of the 
WFPC2 images. NICMOS 
Camera 3 uses a 256$\times$256 HgCdTe array with plate scales of 
0.203859 and 0.203113$\arcsec$ per pixel in x and y, providing a $\sim 52.19\arcsec\ \times 52.00\arcsec$ field of view
(Thompson \etal 1998).  The F187N and F190N filters with effective wavelengths of 1.87 and 1.90 $\mu$m were used to obtain on and off-band images. The FWHM resolution 
is 0.19$\arcsec$ at 1.87 $\mu$m.  Observations at each of the 9 mosaic positions were done using a square dither in each filter setting and at each dither position,
non-destructive reads (MULTIACCUM) were taken.  The total
integration times for each filter are listed in Table 1. 

 The data were reduced and calibrated using the CALNICA version 3.3
task (Bushouse \& Stobie 1998) in IRAF\footnote{IRAF is distributed by the National Optical Astronomy Observatories, which are operated by the Association of Universities for Research in Astronomy, Inc., under cooperative agreement with the National Science Foundation.}/STSDAS and the reference
files (static data quality, detector read noise, detector
 non-linearities files) from the Space Telescope Science Institute
(STScI) NICMOS pipeline, with the exception of the flat-field, and
 dark frame corrections that were provided by the NICMOS Instrument
 Development Team. The dithered images were then shifted and mosaiced
 using the NICMOSAIC and NICSTIKUM IRAF tasks (D. Lytle 1998, private
 communication). The plate scales of
the final ''drizzled'' images are 0.0381$\arcsec$ and 0.0378$\arcsec$ per
pixel in $x$ and $y$. The images were mosaiced with relative offsets determined from common stars in the overlap regions and rotated with north up and east to the left using the data-header orientation angle (see \cite{pol01}).

\subsection {WFPC2 and NICMOS Flux Calibration}

Flux calibration of the WFPC2 images is done as described in Polletta
\etal (2001). For the NIC3 images, we employ scale factors of
 5.050$\times10^{-5}$ and  5.033$\times10^{-5} $
Jy~(ADU/sec)$^{-1}$ at  1.87 (P$\alpha$) and 1.90 $\mu$m (\cite{rie01}).  The rms noise in the final images
is typically 62 and 59 $\mu$Jy (arcsec)$^{-2}$ at
1.87 and 1.90 $\mu$m. As a check on the flux calibration, three of the 
coauthors of this paper independently calibrated the H$\alpha$ and P$\alpha$
images with similar results (within 10\%).

At zero redshift, the H$\alpha$ filter on WFPC2 transmits the [NII] lines (6548.1 and 6583.4\AA
~in addition to H$\alpha$ and stellar continuum. At the redshift of M51 (Z = 0.00154),
the transmissions are 98.7, 0.0 and 91.2 \% for the two redshifted [NII] lines
and H$\alpha$, respectively. Adopting a total flux in the [NII] lines
of 0.4 $\times$ H$\alpha$ in M51 and a flux ratio of 1:3 for the two [NII] lines, we find that the detected line flux will be 1.012 times that of H$\alpha$. 
This fortuitously happens because the redshift of reduces transmitted 
flux of H$\alpha$ by nearly the same amount that the transmitted flux of 
the  6548.1\AA [NII] line is increased. We therefore neglect this 
1\% correction.   

\section{Images}

 The mosaic of continuum-subtracted H$\alpha$ images is shown in
Fig.~\ref{halpha_large} for the central 281$\times$223\arcsec\ . ~The
H$\alpha$ is displayed in red, the continuum B band in blue and the I or Y bands in green. H$\alpha$ emission extends out 10\arcsec\ from nucleus
at PA $\sim$ $-$15\deg and bright, discrete H$\alpha$ emission regions outline
the spiral arms.  The locations of bright H$\alpha$ emission are closely
associated with the dark dust lanes, but relative to the dust (and the mm-CO
emission, see \cite{pol01}), the H$\alpha$ is often displaced to the outside
or leading edge of the arms. In the standard picture of spiral pattern
streaming shown in Fig.~\ref{spiral}, this offset implies that the HII
regions develop subsequent to the time of maximum concentration of the dust
and molecular clouds. Close inspection of Fig.~\ref{halpha_large} also
reveals a number of smaller dust lanes and HII regions in the interarm
regions both to the east and west of the nucleus. 

Two very luminous
associations of stars are also seen --- one approximately 32 \arcsec\ NE of
the nucleus with surrounding H$\alpha$ emission, the other 99 \arcsec W of
the nucleus, just outside the spiral arm. The latter region is remarkable in
being very large in extent ($\sim$ 7 \arcsec\ or 326 pc in diameter) and in having
very little H$\alpha$ emission. The continuum colors and the lack of
H$\alpha$ suggest that this is an aging association in which the O-stars
have mostly evolved off the main sequence ($\geq 10^7$ yr, see
\cite{pol01}). The nature of the larger association is not at all clear since
it is probably much more massive and more extended than any of the younger clusters required to 
power the H$\alpha$ emission regions.

The NICMOS P$\alpha$ (red) image is shown in Fig.~\ref{palpha_large} overlayed on the 
V (green) and B (blue) continuum. In Fig.~\ref{halpha_palpha}, the 
P$\alpha$ (green) is combined with the H$\alpha$ (red) in order to highlight those
regions with relatively high P$\alpha$/H$\alpha$ ratios, indicating particularly 
high dust reddening. For a smaller area west of the nucleus, Fig.~\ref{halpha_palpha_small} 
shows the H$\alpha$ and P$\alpha$ emission with no optical continuum background. Although generally P$\alpha$ shows the same emission regions
as H$\alpha$, it is clear even from visual inspection of Fig.~\ref{halpha_palpha} and ~\ref{halpha_palpha_small}
that many of the arm HII regions have considerable reddening.  There are  several regions 
appearing green in Fig.~\ref{halpha_palpha}, implying strong P$\alpha$ but only very weak H$\alpha$. Many
HII regions also exhibit strong gradients in the P$\alpha$/H$\alpha$ ratio.

\section{HII Region Parameters}

In measuring HII regions, the high angular resolution HST is critical 
to resolving regions associated with individual OB star clusters. HST-NICMOS 
also enables  measurement of P$\alpha$ (which can't generally be done from the 
ground); in combination with H$\alpha$, this line provides a probe of the dust extinction. These points are illustrated well
in Fig.~\ref{halpha_comparison} which  shows a region approximately 60\arcsec\
N of the nucleus  at the full resolution H$\alpha$ and the H$\alpha$ smoothed to 1.5\arcsec\ resolution (to simulate ground-based imaging). Comparison of two the two H$\alpha$ images underscores the need 
for high angular resolution in order to resolve the multiple, distinct 
HII regions which often exist in a single complex.

In Fig.~\ref{sub_area},  the H$\alpha$ and P$\alpha$ are shown for the same region and large variations can be seen in their relative distributions.
due to variable reddening. For example, the southern end of the complex is strongest in P$\alpha$
while the northern region is brightest in H$\alpha$. 

\subsection{HII Region Definition and Measurement}

Measured HII region properties will be fundamentally dependent on both the resolution of the images and 
the procedures used to define the boundaries of the emission regions. For example, higher resolution images will resolve some of the 
larger regions into multiple components (and hence reduce the number
of the most luminous regions). In addition, the criteria adopted to resolve (or break up) blended regions can be critical. Algorithms 
to separate out the regions from the 
background and break up blended regions might invoke logic 
based on {\it a priori} understanding of the characteristics 
of Galactic HII regions.  For example, virtually all Galactic 
HII regions have diameters much less than 50 pc so it is reasonable to 
expect that extragalactic regions larger than this are likely
to be blends, ionized by multiple OB star clusters. Lastly, 
data with higher sensitivity 
might join multiple regions which would have appeared
separate at lesser sensitivity if the threshold for joining regions was set at a sufficiently 
low level.

The HII region measurement process may be broken into three steps :
1) the definition of the boundaries of emission regions at a
specified intensity threshold, 2) the resolution of multiple
strong peaks within a single boundary into separate, discrete emission regions
and 3) measurement of HII region parameters such as peak and integrated
fluxes, sizes and positions. Clearly, the breaking up 
of blended features is the most difficult (and subjective) step  ---  for this reason we developed an automated procedure (rather than an interactive process). This has the advantage that the results can be carefully compared
for different input parameters and the selection algorithm is  
uniform across the entire image. Extensive 
trials of the selection parameters were used to test their effects on the results.  

Prior to measurements of the {\it discrete} HII regions, we removed a very extended background in order 
to 'partially' remove the diffuse ionized gas (DIG) emission and to suppress any residual continuum which might result from color 
variations of the continuum in different regions of the galaxy.  This background at each pixel was estimated from the 30'th percentile intensity for a histogram of pixels in an area of 64 $\times$ 64 pixels
centered on each pixel (that is 70\% of the pixels are brighter than the subtracted background). The local histograms of pixel values in the final background-subtrated image
have a large number of 
pixels close to zero and a small fraction on the positive tail, representing real H$\alpha$ emission
(assuming the HII regions don't fill a large fraction of the area). Sixty-four pixels corresponds to approximately 300
pc linear scale and this is much larger than any discrete HII regions.  The 30'th percentile was chosen to avoid significant Malmquist bias and to avoid 
removal of discrete HII region line emission in areas with large numbers of HII regions (none of which fill 
70\% of the pixels in a surrounding 64 $\times$ 64 pixel box).  After removing the background, the noise ($\sigma$) was estimated in areas not including obvious emission. Typically, $\sigma \simeq 4.8\times 10^{-18}$ erg cm$^{-2}$ s$^{-1}$.  The removed background varies between between -0.9 and 2.8$\sigma$ and the average values 
are -0.04 (whole image), 0.08 (arms), 1.8 (nucleus) and -0.15 (interarms) $\times \sigma$ (as quoted above). Discrete, but low surface brightness 
HII regions would still be catalogued, provided they have at least one pixel exceeding 6$\sigma$ 
(see below).

The logic adopted for definition of the boundaries of the emission regions was : start from the brightest pixel, find all pixels connected to this peak
down to a level of 65\% of the peak value; then repeat the procedure 
starting from the brightest remaining pixel; and so on until there are 
no pixels left above the adopted threshold for a 'significant' peak (here taken to be 6$\sigma$). The only complication is that as one is collecting 
the neighboring pixels around a peak, one must also check that those 
new pixels wouldn't more appropriately associate with one of the peaks 
found earlier in the process. In such instances where there were multiple
regions neighboring on a pixel, the pixel was attached to the region
which had the highest average value for its neighboring pixels.

that it is essentially assuming a constant 'curve of growth' as a function 
the percentage-of-peak for all HII regions, i.e. the HII regions must 
all have homologous morphology.

In detail, our procedure involved the following steps :

1) Find the peak pixel remaining in the image. 

2) Let this pixel be the basis for starting a new "peak".

3) Generate a list of all pixels above 65\% of this peak value. 

4) For each of the pixels found in \# 3, associate it with the new peak or whichever pre-existing 
peak has the highest average in pixels which touch the pixel in question. As pixels
are assigned to HII regions, they are also removed from the original
image to avoid further consideration. Any of the pixels in the 
list from \# 3, not touching on a previous HII region, are left in the image for later
consideration. Thus one is gradually working
down in intensity through the image. 

5) If there are remaining pixels above 6$\sigma$ which could form the basis
for a new peak, then go back to step \# 1.

6) Lastly, the algorithm examined all the HII regions defined in steps 1--5
to see if any should be collected into a single region. Any region which did not have a valley of at least a 4.5$\sigma$ (down 
from its peak value) between it and the border pixels of any neighboring 
HII region was added into the neighboring region.   

For any regions with more than 3 pixels, a curve of 
empirical growth was derived to correct for flux outlying pixels beyond the lowest non-zero
pixel value. 
The curve  of growth was computed {\it individually} for each HII region,
from measurements of the total integrated flux within each HII region as a function of the pixel intensity, starting from the peak value and working down to 2.5$\sigma$. For example, including all pixels in the region with pixel values 
above 2.5, 4, 6, 8, 10 ... $\sigma$, the separate flux sums were calculated. These flux integral measurements were then fit by a straight line (as a function of the variable cutoff threshold) and 
the fit was extrapolated to 0 intensity cutoff value. This correction
resulted in a median increase of only 15\% in the flux. In this manner,
we are not assuming a fixed morphology for all HII regions and the 
extrapolations are extremely modest since we have collected all connected pixels down to only 2.5$\sigma$.

Although we are using a criterion based on a percentage-of-peak for the {\it initial} association of the nearby pixels with a given peak pixel, this technique is not the same as that developed by McCall \etal (1996). They collected pixels down to a fixed percentage of the local peak but then extrapolated the flux within
this region using a fixed multiplicative factor to estimate the total flux of the HII region. Here, the 65\% is simply used as the initial 
criteria for associating neighboring pixels with a particular local maximum.
And as we continue working down in intensity, pixels below the 65\% level
(all the way down to 2.5$\sigma$) are associated with the pre-existing nearby peaks. The technique employed by McCall \etal (1996) suffers the obvious problem 
Actually, our algorithm is somewhat similar in spirit to that developed by Thilker (2000). Both algorithms, let all HII regions grow simultaneously as one works down to lower flux levels in an image (see step 3 above). This ensures 
that as much flux as possible is gathered into each HII region without
bias towards the order with which each HII region was first found. In our case,
new HII regions are always started from the highest remaining pixel in the 
entire image and we require that every distinct HII region must have a least 
a 4.5$\sigma$ valley between its peak and any pixels in a neighboring 
HII region (see step 6 above). 

In the course of refining the measurement technique several of the
parameters were experimented with and the results compared with what we
 would have selected by 'eye' as distinct HII regions. For example, the
percentage of the peak flux above which pixels can be added and the
level below which the image was truncated (finally adopted at
 2.5$\sigma$) were varied. The results were not extremely sensitive to the 
precise values of these parameters. 
The sample area north of the nucleus is shown in contour form for H$\alpha$ in the upper left panel of Fig.~\ref{sub_area} and the derived features are illustrated in in the lower right panel of Fig.~\ref{sub_area}.

The algorithm was also tested on a simulation image in which spherical 
HII regions with a uniform distribution of Lyman continuum 
output (with a range of 10$^2$) were positioned randomly within a 3-d spatial cube. The regions were postulated to have identical electron densities
so as the the Lyman continuum increased, the HII region diameter
would be larger. The cube was then projected 
on two dimensions and processed with the HII region finding algorithm. 
The HII regions found agreed very closely with the input HII 
region locations and sizes. The simulation and measurement was run 
for increasing numbers of input HII regions ranging from 10 to 100
in order to test the effects of blending. (Given the input sizes of the HII
regions and the size of the 3-d cube, the case with 150 regions corresponded to very severe blending.) In all instances, the 
HII region finding algorithm found the correct number and parameters for 
the input regions (to within $\leq$ 20\%). Test fits for the 
dependence of the total flux on the measured 
diameter was typically L$\propto D^{2.75}$, compared to the 
theoretical D$^{3}$ dependence -- probably due to limited 
spatial resolution. The fact that the results came so close to 
recovering the input parameters was particularly reassuring in 
the case of the 100 region simulation since the images were heavily 
blended in this case.  

Once the boundaries of the H$\alpha$ emission regions were defined,
the peak, the integrated flux, the 
emission centroid coordinates and number of pixels within each region were catalogued. The size of each region was calculated as (area)$^{1/2}$. 

The final list has 1373 H$\alpha$ emission regions : 
1321 having at least 3 pixels and 1273 having 5 or more pixels. The largest region associated with  the nuclear AGN jet
had 895 pixels was excluded from the following analysis. The total flux
of the 1373 discrete HII regions constitutes 31\% of the total H$\alpha$ flux from the 
area of the galaxy covered in the WFPC2 images; the remaining flux constitutes 
the DIG emission, discrete HII regions with peak flux below our 6$\sigma$
peak requirement and possibly some unremoved continuum. Given the low surface brightness of the residual flux (and its low SNR)
we do not feel the data enable an investigation of the source of the 
DIG emission -- specifically whether it is truly diffuse or simply
discrete, but low surface brightness HII regions.  

As a check of the HII region definition and measurement routines,
we compared our results for selected individual regions with ground-based 
H$\alpha$ measurements (\cite{ran92}, \cite{pet96}, \cite{thi00}). 
Although Rand (1992) did not publish his individual HII region measurements,
he did provide us with his HII region list in tabular form.  Fig.~\ref{hii_sp_dist_r92} shows the locations of the H$\alpha$ emission regions from  Rand's and our 
samples. Generally, good agreement can be seen in the 
areas of common coverage (i.e. excepting the nucleus which was not measured by Rand and the outer disk which was not covered in our images). 
However, in most cases, the higher resolution HST data 
resolves multiple HII regions within individual regions identified 
in the ground-based imaging. This 
is illustrated in Fig.~\ref{halpha_comparison} which shows the HST data 
in a small area north of the nucleus at the original HST resolution 
and smoothed to 1.5\arcsec\ resolution to simulate ground-based imaging.
In the area covered in our study, there were 17 regions catalogued by Rand 
which were not catalogued by us. We 
therefore also compared with the HII region catalogue of Petit \etal (1996)
and in most cases the regions catalogued by Rand but not 
seen by us were also not present in Petit \etal (1996); however, the Petit
survey had lower sensitivity than Rand's. As an additional check,
we also examined on a case by case basis the 17 Rand regions which we did not 
find. For one, we saw no evidence of a source in either the 
H$\alpha$ or the H$\alpha$+continuum image; one was on the edge of our field; 
and the remaining regions had peaks between 3.4 and 5.6$\sigma$ in our images.
Thus the latter sources did not meet our 6$\sigma$ criterion for the peak flux. In fact, the excellent correspondence between our sample and Rand's is 
quite impressive given the fact that our resolution is aproximately 100 times smaller
in area and therefore an extended, low surface brightness feature in Rand's image could easily be missed here. 

 Comparison of our fluxes with Rand's was not straightforward since
he did not remove the large-scale diffuse background but rather a local 
background (immediately around each region). Nevertheless, the results appear consistent :  for 10 regions
with well-defined, bright H$\alpha$ emission, the integrated fluxes were
found to agree within 25\% in all cases. The peak fluxes and 
sizes are of course resolution
dependent and perfect agreement should not be expected.
Thilker \etal (2000) 
also compared their results with Rand's (1992) measurements and found good agreement. 

Thilker developed an iterative grouping algorithm for delineation
of HII region structures and he kindly made his routines available. In the end,
we developed our own algorithm for reasons of simplicity,
familiarity with our routine and speed of execution. Nevertheless, we
did test the two procedures in the area of the northern spiral arm shown in 
Figs.~\ref{halpha_comparison};
the regions which were defined were consistent but not identical.

Our algorithm was also run on the P$\alpha$ image (using the same parameters
specified in terms of the image noise level), yielding 232 P$\alpha$ regions.
All except three of these regions were inside one of the previously found H$\alpha$ regions; however, in most cases the boundaries 
were somewhat different. Usually, the P$\alpha$ region was smaller in size, due to the intrinsically lower flux and SNR of P$\alpha$. In a few instances, the peak of the P$\alpha$ was very 
significantly shifted from the H$\alpha$ peak and it appears that
these are regions with particularly high extinction at the 
location of maximum P$\alpha$ emission. 

\section{Clustering of HII Regions}

In order to quantify the effects of HST versus ground-based image resolution and to illustrate the clustering of the HII regions measured here, we 
have computed the two-point angular correlation function for the measured 
centroid positions of the 1373 HII regions (top panel of Fig.~\ref{two_pt}). 
In the lower panel of Fig.~\ref{two_pt}, the correlation function 
is shown for {\it pixels} with H$\alpha$ emission exceeding 3$\sigma$ (in the 
same  H$\alpha$ image with background subtraction which was used for defining and measuring the discrete HII regions. In both instances,
the angular sampling of the images was normalized by the 
correlation function for pixels sampled randomly
within the observed region. However, due to the linear association
of HII regions along the spiral arms and the fact that their number is 
generally increasing towards small galactic radii, the purely random normalization does not
remove large-scale correlations -- thus neither correlation function 
goes asymptotically to 0 at large angles nor does the integral equal 0. 

{\it Both} correlation functions clearly show a strong clustering inside 2\arcsec,
corresponding to 92 pc.
This is inside the spatial scale sampled in ground-based H$\alpha$ images,
indicating that a large number of the regions classified from ground-based imaging are resolved into multiple regions at the 0.1\arcsec\ resolution used
here. Pleuss \etal (2000) reached a similar conclusion based on a Minimal Spanning Tree analysis of HST images for M101. 

The correlation function for the discrete HII regions (top panel) 
shows a decrease inside 1\arcsec which is a result of the fact that 
the algorithm used to define the HII regions necessarily requires {\it at least}
one pixel lower by 40\% between any two HII regions in order to subdivide an 
H$\alpha$ emission region into two regions. 

In Fig.~\ref{nearest} the percentage of HII regions with nearest neighbors 
with separation less than $\theta$ is shown as a function of $\theta$. 47\% have their nearest neighbor within 1\arcsec\ and 81\% within 2\arcsec. Only 10\%
have their closest nearby HII region more than 4\arcsec\ away. Both the correlation
function and  Fig.~\ref{nearest} clearly show that the HII regions are 
strongly correlated on scales less than 2\arcsec.

\section{Luminosity Functions}

The {\it observed} (uncorrected for extinction) H$\alpha$ luminosity function  is shown in Fig.~\ref{lf_comparison_same}
together with those derived by Rand (1992) and Petit \etal (1996), using only HII regions from their studies which are in the same area of the galaxy covered by us. The two ground-based luminosity functions extend at least a factor of 5  higher
in luminosity. This is due to the fact that all of the most luminous regions
catalogued in the ground-based studies were either resolved into multiple 
regions in our sample. 

To characterize
the luminosity function, we express it as a truncated power
law : in differential form,  

$$ {d~N(L_{H\alpha}) \over d~ln~L_{H\alpha}} =  N_{up} \left( L_{H\alpha}  \over {L_{up}}\right)^{\alpha}
 ~~ \eqno (1) $$
 
\noindent and in integral form,

$$ N(\ge L_{H\alpha}) =  {N_{up} \over -\alpha} \left[\left(L_{H\alpha} \over {L_{up}}\right)^{\alpha} - 1\right]
 ~~ \eqno (2) $$

\noindent (\cite{mck97}). (Note that we adopt the opposite sign 
convention for $\alpha$ from that used by \cite{mck97}.)  As pointed out by McKee \& Williams (1997), this 
form has the advantage of having clear physical interpretations for the 
parameters. Specifically, L$_{up}$ is the highest luminosity region, N$_{up}$
is approximately the number of regions between 0.5 L$_{up}$ and L$_{up}$ 
for $\alpha \sim$ -1 and 
N$_{up}$/$-\alpha$ is the number of regions expected above L$_{up}$ if the 
distribution were not truncated (i.e. one can directly see if the distribution
is 'significantly' truncated or terminated by low number statistics).

Fitting Eq. 2  
to the apparent luminosity function yields $\alpha$ = $-$1.01$\pm$0.04, compared to $-$0.50 and $-$0.32 derived for the Rand (1992) and Petit \etal (1996) samples over the range 
$40$ --- 800$\times 10^{36}$ erg s$^{-1}$. (Since these fits are
for source counts binned logarithmically in L$_{H\alpha}$, the exponents should 
be decreased by 1 for N(L)dL.) Our power-law index is substantially steeper than that derived from previous
studies (\cite{ken89}, \cite{ran92}, \cite{pet96} \& \cite{thi00}) due to the fact that the higher angular resolution resolves the more luminous, apparently blended regions. The slope derived here is similar to that of Galactic radio HII 
regions ($\alpha$ = $-$1 to $-$1.3; see \Section{12}). 

In addition
to a steeper slope, we also find that the {\it observed} luminosity function
doesn't extend to as high luminosity as the ground-based LFs due to our resolving the larger regions. However, the lower maximum-{\it observed} luminosity  
obtained here is compensated 
by high extinction corrections (see \Section{10}).
The mean extinction derived in \Section{10.2} for the discrete H$\alpha$ emission regions is A$_{H\alpha}$
= 0.798$\times<$A$_V$$>$ = 0.798$\times$3.1 = 2.47 mag. If this is applied  on-average to all HII regions, then the observed H$\alpha$ luminosity functions 
are increased by a factor of $\sim 10$ (\Section{10.2}). 

The {\it observed} luminosity functions differentiated between arm, interarm and nuclear areas are shown in Fig.~\ref{sep_lf_4}. These three areas are defined and illustrated in Polletta \etal (2001). Over the luminosity range L$_{H\alpha}$ =
$12 \to 500\times10^{36}$ ergs s$^{-1}$, logarithmic truncated power-law fits have exponents
$-$0.72$\pm$0.03, $-$0.95$\pm$0.05, and $-$1.12$\pm$0.11 for the arm, interarm and nuclear (excluding the nuclear jet) regions, respectively (see Table 2). Thus, the luminosity function is significantly 
flatter in the spiral arms than in the interarm regions. 
Rand (1992) and Thilker \etal (2000) also found steeper luminosity functions in the interarm
regions (exponent $-$0.93 $\to$ $-$0.96 versus $-$0.48 $\to$ $-$0.72); however, the actual values of their slopes in the interarm and arm areas are somewhat different from ours.

There are several possible explanations for the flatter luminosity
functions in the arms and the nucleus compared to the interarm regions :
1) the surface density of HII regions is higher in the 
arms, causing more blending/clustering which in turn produces a large number of 
apparently high luminosity regions; 2) the OB star cluster mass functions are significantly different in the two areas; and 3) the interarm
HII regions and associated OB star clusters are, on average, older and therefore
have evolved to lower Ly continuum output levels. The only way there could be
a systematically older population of clusters in the interarm
regions than in the arms would be if the OB star clusters in the disk formed within the arms
and then aged as they moved into the interarm regions. However, the 
duration of the Lyman continuum emission from a coeval cluster is far too short ($\leq 3\times10^6$ yr; see \Section{13.2}) compared to the 
time ($\geq 3\times10^7$ yr) needed to migrate into the interarm regions for this explanation 
to be viable. Although the second 
explanation can not be ruled out at present, we think that the 
level of blending/clustering expected in the spiral arms
is entirely consistent with the notion that many of the most luminous 
regions are, in fact, blends. One supporting piece of evidence for 
this is the clear trend for the most luminous regions to be the 
most extended and to have the lowest densities (see \Section{9}).

The fitting results summarized in Table 2 clearly show a {\it significant} truncation of the lumninosity functions at L$_{up} \simeq 4\times 10^{38}$ erg s$^{-1}$. The last columns of Table 2 give the fitted N$_{up}$ and N$_{obs}(L>L_{up})$. As noted in the discussion following Eq. 2, the truncation is significant if N$_{up}$/$-\alpha$ $>>$ N$_{obs}(L>L_{up})$. This is 
indeed true for all four fits (based on comparison of the last two columns in Table 2), strongly suggesting a physical or observational limitation to 
the maximum luminosity of the HII regions.

In none of the luminosity functions do we see evidence of the break in the slope at L$_{H\alpha}$ $\sim$
10$^{38.6}$ ergs s$^{-1}$ which has been reported in some ground-based studies
(\cite{ken89} and \cite{ran92}). (We did experiment with smoothing 
our H$\alpha$ images by a factor of 4, then redefining the boundaries and determining
the luminosity function. The result was that the derived luminosity function became shallower and developed a spectral break; however, since the 
break was somehwat lower in luminosity than 10$^{38.6}$ ergs s$^{-1}$ and we can not say that it is a result of blending.)  

\section{Lyman Continuum Emission Rate}

The observed HII regions range in luminosity from L$_{H\alpha}$ = 2$\times 10^{36}$ --- 2$\times 10^{39}$ erg ~s$^{-1}$. For case B recombination the H$\alpha$ luminosity is  

 $$ L_{H\alpha}  = 3.55\times10^{-25}\left({T_e \over 10^{4}K } \right)^{-0.91} n_e n_p V ~{\rm erg~s^{-1}}  \eqno (3) $$

\noindent where V is the volume of the HII region and n$_e$ and n$_p$ are the electron and proton volume densities (Osterbrock 1989). Since absorptions by He do not significantly reduce the number of photons available to 
ionize H (Osterbrock 1989), the required Lyman continuum production rate, Q$_{LyC}$, is then given by  

 $$ Q_{LyC}  = 7.32\times10^{11} L_{H\alpha} \left({T_e \over 10^{4}K } \right)^{0.11}{\rm s^{-1}} . \eqno (4) $$

\noindent The {\it observed} H$\alpha$ luminosities therefore translate to 
Q$_{LyC}$ = $2.2\times 10^{48}$  ---  $1.5\times 10^{51}$ s$^{-1}$ for T$_e$ = 10$^4$ K.
Extinction corrections are discussed in \Section{10}.

\section{HII Region Sizes and Densities}

The HII region luminosities and their Lyman continuum emission rates constrain the implied 
masses of the OB star clusters; similarly the sizes and electron densities
of the HII regions reflect on the evolution of their Str$\ddot{o}$mgren spheres
and the surrounding ISM. 

\subsection{Sizes}

The distribution of HII region sizes (D = 2$\times$(area/$\pi$)$^{1/2}$) is shown in Fig.~\ref{histo_sizes} for all regions with at least 2 pixels. They range 
from 10 to 250 pc in diameter;  those $\ge$ 120 pc were not plotted since they are clearly blends. The lower limit of 10 pc 
is due to the minimum 2 pixel criterion. The mean sizes are 30 --- 34
pc in arm, interarm and nuclear regions (Fig.~\ref{histo_sizes}). Both the shapes of the distributions and the mean sizes are
similar for all three areas. The majority of the HII regions have size
$\leq$ 50 pc and thus could be associated with a single or a few
OB star cluster(s). Those regions with larger sizes are {\it very} likely blended 
superpositions of multiple (but possibly related) OB associations (see \Section{11}).

\subsection{Electron Densities (n$_e$)}

Since the H$\alpha$ luminosities
are proportional to the volume-integrated emission measures, the mean electron density can be obtained using the measured sizes :

$$ <n_e>   = 43 \left({ ~(L_{H\alpha~cor}/10^{37}~erg~s^{-1})~~(T/10^4 K)^{0.91} \over (D/10~pc)^3 }\right)^{1/2} 
 ~~ cm^{-3} . \eqno (5) $$

\noindent In Eq. 5, we normalized 
to typical values of L$_{H\alpha~cor}$ = 10$^{37}$ ergs s$^{-1}$ and a size of 10 pc. 
The derived $<n_e>$ for the HII regions are distributed 
mostly between 5 and 20 cm$^{-3}$. In Fig.~\ref{lum_size_density} the HII 
region luminosities and sizes are plotted as functions of mean electron density. The sharp boundaries to the distributions on the lower side are 
 due to the surface brightness threshold and the minimum 
size of 1 pixel within the HII region. The average
value is $<n_e>\simeq$ 13 cm$^{-3}$ for sizes less than 40 pc. If the apparent n$_e$ are corrected for the average extinction correction (Eq. 8), the mean
densities are increased by a factor of 3 to $<n_e>\simeq$ 39 cm$^{-3}$ which is similar
to that of resolved Galactic HII regions ($n_e \sim$ 100 cm$^{-3}$). 

In the above estimates we have made no corrections 
for the fact that many of the HII regions are only marginally resolved; such
corrections would of course increase the mean densities further. Some of the larger regions undoubtedly do include substantial empty or neutral volumes. 
Moreover, there are substantial gradients in the H$\alpha$ surface brightness
within individual regions, indicating that the internal densities 
are non-uniform and commonly increase several-fold at the peaks. 

\subsection{Luminosity vs Size : HII Region Blending}

In Fig.~\ref{lum_size}, the HII region luminosities are plotted as a function of their measured sizes. There is a clear trend for the more luminous regions to be more spatially. Although 
such a trend is certainly expected, since more luminous OB associations
can ionize a larger volume of gas, the observed dependence is shallower
than the $D^{3}$ dependence expected if the electron densities were constant. 
Specifically, the more luminous and larger regions appear to have lower average n$_e$. The luminosity versus size data is fit well by L $\propto D^{2}$. (The lower limit of the data points in Fig.~\ref{lum_size} is 
caused by the observational threshold on the surface brightness.
However, this threshold can not account for the decrease in density
for larger regions -- if the larger regions had the same densities as the 
smaller regions, they would more easily make it into our sample.) The proper explanation for the correlation is almost certainly that the larger regions
are blended and they include substantial empty or neutral volumes
or equivalently, that the large HII regions may have expanded outside their progenitor GMCs. 

In fact, it is easily shown that for superposed or blended HII regions
one expects precisely the L $\propto D^{2}$ correlation which is 
observed. If the boundaries of 'N' multiple regions of the same 
size just overlap as projected 
on the plane of the sky, their total projected 'size' (D$_T$), computed as the square root of the sum of the total area, is given by D$_T^2 = N\times$ D$_i^2$. Since the total luminosity of the blended region is L$_T = N L_i$, then
we expect L$_T = D^2\times L_i/D_i^2$. This result is obviously not dependent
on the HII regions being identical; {\it the observed correlation between
luminosity and size is thus strong evidence that the high luminosity 
regions are blends or superpositions of multiple lower luminosity regions}. 
This is certainly not surprising based on observations of Galactic 
HII regions which often showed multiple centers of ionization 
on scales of less than a few parsec (see \Section{12}).  
   
\section{HII Region Extinctions from H$\alpha$/P$\alpha$}

For case B recombination in an ionization-bounded HII region, the intrinsic 
flux ratio for H$\alpha$/P$\alpha$ is 8.15 and the non-Lyman series lines should be optically thin (\cite{ost89}). (For a density-bounded HII region, the case A recombination line ratio is H$\alpha$/P$\alpha$ = 6.14 but Case B is generally 
adopted as more appropriate for high surface brightness, discrete HII regions.) The observed flux ratios can be less than 8.15 due to the higher extinction 
at the $\lambda$ = 6563\AA~(H$\alpha$) compared to 1.87$\mu$m (P$\alpha$). 
We have used the observed ratios to estimate the mean extinctions 
of each HII region detected in both lines, using the relation

  $$ A_{V}  = 3.75\times log \left({ 8.15~F_{P\alpha} \over F_{H\alpha} }\right) 
~{mag} . \eqno (6) $$

\noindent The constant in Eq. 3 was derived assuming the standard Galactic extinction 
curve (\cite{rie85}, \cite{car89}) and assuming the dust is distributed in a foreground screen, uniformly covering each pixel. 

After smoothing the H$\alpha$ to the same resolution as P$\alpha$, we measured the  H$\alpha$/P$\alpha$ flux ratios and extinctions at all pixels within the HII 
region boundaries using only pixels which had both H$\alpha$ and P$\alpha$ detected at $\geq 5\sigma$. 209 HII regions met these criteria. (If the H$\alpha$
is smoothed to the P$\alpha$ resolution, the HII region finding algorithm found  702 regions. However, for the extinction analysis here, we adopted the original high resolution HII region definitions simply because they yielded better boundary definition.)
Flux ratios and extinctions were then estimated for each HII region
by averaging the pixel- ratios and extinctions, rather than from ratios of the  HII region-integrated fluxes. The distribution of these average flux ratios and A$_V$s for the 209 regions is shown in Fig.~\ref{hii_ratio_ext}. The mean extinctions are : A$_{V}$ = 3.1 mag
(weighting each region equally), 2.4 mag (weighting each by the 
observed HII region luminosity) and 3.0 mag (weighting by the extinction-corrected luminosity).  The derived mean 
extinctions are in reasonable agreement with 
the values (A$_V$ = 0.8 --- 4 mag) obtained by van der Hulst \etal (1988) for a sample 37 HII regions in M51 comparing H$\alpha$ and radio free-free fluxes at 8\arcsec\ resolution. For 14 of the regions, the Balmer decrements yielded considerably lower values (in the range A$_V$ = 0.4 --- 2.4 mag; \cite{van88}) but these are optically biased towards lower extinction regions. It should be pointed out that since the HII region boundaries 
were defined from the H$\alpha$ images, the analysis above excludes regions
of extremely high extinction in which the P$\alpha$ emission is 
significantly offset from H$\alpha$. The derived 
extinctions would have very likely been higher if the P$\alpha$
had been used to define the HII regions. We used H$\alpha$ 
for delineating the HII region boundaries simply because of the higher SNR and resolution. 

In Fig.~\ref{av_lha}, the derived extinctions of the 209 regions are plotted against 
the {\it observed} (not extinction-corrected) H$\alpha$ luminosities. In this figure,
the filled circles are the average values in separate luminosity bins.  No strong correlation (or anti-correlation) is seen between the 
{\it observed} luminosities and the derived extinctions as might be expected if the flux variations were largely due to extinction variations. The filled-circle  averages do  
show a weak correlation in the sense that the fainter regions
have somewhat higher extinction, but the dispersion within bins 
is clearly much larger than the trend. 

We also measured the flux ratios and derived extinctions for all pixels in the
H$\alpha$ and P$\alpha$ images which were both detected at $\geq 5\sigma$, independent
of whether or not they were inside one of the discrete H$\alpha$ regions. The 
results (shown in Fig.~\ref{Pa_vs_Ha_pxl_th5}) are entirely 
consistent with those found within the discrete HII regions 
(Fig.~\ref{hii_ratio_ext}). When the selection threshold was increased to $10\sigma$, the results were also similar, implying  
that the larger extinction values are not simply the result of 
poor signal-to-noise or a bias in either image. 
 
\subsection{Extinction Gradients Across HII Regions}

Detailed comparison of the H$\alpha$ and P$\alpha$ images reveals 
that most individual HII regions also have very large variations in 
the H$\alpha$/P$\alpha$ ratio across their areas. (It is for this 
reason that the ratios and extinctions for each region were computed 
in \Section{8.0} as averages of the values from individual pixels rather than 
from ratios of the integrated fluxes.) In Fig.~\ref{halpha_palpha_small} the area 
to the west of the nucleus is shown with H$\alpha$ in red and P$\alpha$ in 
green (with no continuum added). Several of the emission regions show entirely different morphology
and peak locations in the optical and near infrared lines. The
gradients and peak offsets vary from region to region; they therefore aren't due to misalignment
of the images. In fact, there are several examples in this image 
of emission regions in P$\alpha$ which are hardly detected 
in H$\alpha$, indicating extinctions in excess of 4 mag. There
are also many regions which are seen in H$\alpha$ but not
P$\alpha$ but this is simply due to the intrinsically high flux of H$\alpha$ together with the lower sensitivity of the P$\alpha$ image.
 
\subsection {Extinction-Corrected HII Region Luminosities}

The extinction-corrected H$\alpha$ luminosity is obtained 
from  

  $$ L_{H\alpha~cor}  = 10^{0.320 \times A_{V}} \times L_{H\alpha~uncor} 
 , \eqno (7) $$

\noindent using the standard ISM extinction curve (\cite{rie85}) for which 
A$_{H\alpha}$ = 0.798 A$_V$.
For the mean extinction (A$_V$ = 3.1 mag; \Section{10}), 
the observed luminosities are increased by factors of $\sim$9.9 and we will adopt 
the an average correction factor of

  $$ <f_{H\alpha~ext-cor}>  \simeq 10  \eqno (8) $$

\noindent as an appropriate general extinction correction where individual extinctions are not available.

The {\it extinction-corrected} and {\it observed} luminosity distributions are shown in Fig.~\ref{lf_ext_cor} for the 209 regions detected in both H$\alpha$ and P$\alpha$.  (The 
extinction-corrected luminosity of each region was calculated by correcting each pixel for 
its extinction and then summing the extinction-corrected pixel luminosities.)
The {\it extinction-corrected} distribution exhibits 
a broad peak at approximately L$_{H\alpha}$ = $10^{38}$ erg s$^{-1}$
and the most luminous region is at 3$\times 10^{39}$ erg s$^{-1}$. The falloff
on the low luminosity side is not real since the sample 
has a detection threshold and there are no regions from below the threshold
which can populate the low end after being corrected for extinction. The
distributions shown in Fig.~\ref{lf_ext_cor} should not be interpreted as luminosity functions since they do not include all pixels in each region,
only those pixels detected at $\ge 5\sigma$ in both lines. {\it The critical point we make from Fig.~\ref{lf_ext_cor} is that the net result of extinction 
corrections when they are derived on a case-by-case basis is to shift the luminosity distribution up a factor of 10 in luminosity.}  

\section{HII Region Sample with Diameter $\leq$ 50 pc}

The most luminous emission regions in our sample are quite clearly blends or superpositions of
HII regions with multiple OB star clusters producing the ionization. Their sizes are typically
$> 50$ pc (see Fig.~\ref{lum_size} and \Section{9.1} --- \Section{9.2}). {\it Fifty pc is a very conservative upper limit to the
size of a region which might plausibly be ionized by a single compact OB cluster.} This is because
the MS sequence lifetime is only 3$\times 10^6$ yrs for the OB stars which produce most of the ionizing photons. Within this time, the ionized gas can expand only to radius $\sim 30$ pc even 
if it is {\it freely} expanding at the sound speed (10 \kms) of the 10$^4$ K gas  and after the source of ionization is turned off, the gas recombines on a relatively short timescale ($\leq 10^4$ yrs for n$_e \geq 10$ cm$^{-3}$). Thirty parsec is therefore a very conservative maximum radius for the HII region ionized by 
a single, coeval cluster of stars. It is a 'conservative' maximum since the expansion is usually much slower than 10 \kms even if the surrounding, neutral gas has density $\leq 10$ cm$^{-3}$ (e.g. \cite{ost89}). 

Of course, the early growth of the HII region to its initial Str$\ddot{o}$mgren radius can 
be faster but this initial phase takes place in higher density regions 
and the resulting Str$\ddot{o}$mgren radius is much smaller. The subsequent evolution is hydrodynamic and this is certainly the phase in which we see the observed HII regions. In late phases if the HII region becomes density-bounded,
the ionization front might expand at a higher speed but this would correspond
to the DIG HII regions. 

In Fig.~\ref{lf_50} we show the distribution of observed H$\alpha$ luminosities for 
all 1101 HII regions in our sample with diameters $\le 50$ pc. The range of observed L$_{H\alpha}$ is 2$\times 10^{36}$ to 1$\times 10^{38}$ erg s$^{-1}$. Since the 
derived extinctions are approximately independent of apparent H$\alpha$ luminosity
(Fig.~\ref{av_lha}), we conclude that the maximum {\it intrinsic} luminosity 
of unblended HII regions is 
a factor of 10 higher or 

   $$ Max~single~cluster~L_{H\alpha~cor} \simeq 10^{39} erg s^{-1}  \eqno (9) $$
 
\noindent and the maximum Lyman continuum production is 
        
   $$ Max~single~cluster~Q_{LyC} \simeq 7\times 10^{50} s^{-1}.  \eqno (10) $$

\noindent Once again, this is a conservative upper limit since some of the regions in the $\leq$ 50 pc sample are also likely to be blends.

\section{Comparison with Galactic HII Regions}

For comparison with Galactic HII regions we make use of radio continuum
observations of the free-free continuum to avoid Galactic line-of-sight dust obscuration.
Radio studies have concentrated on 'compact' HII regions
which are relatively young and have high surface brightness. 
Schraml \& Mezger (1969) observed 18 free-free emission complexes  at $\lambda = 2$ cm with 2\arcmin\ resolution (typically corresponding  to 0.5 --- 5 pc) 
and we make use of their sample for our discussion.
Their results are summarized in Table 3 for M43, M42(Orion Nebula), IC1795(W3), W51, and W49 (ordered with increasing luminosity). 

Comparing the implied L$_{H\alpha}$ of the Galactic HII regions with the extinction-corrected H$\alpha$ luminosities (\Section{11}) for the sample with D $\leq$ 50 pc, it can be seen that the first value in the M51 distribution function corresponds to
a few times M42 and the max single cluster L$_{H\alpha~cor}$ (Eq. 9)
corresponds to W49. Since W51 and W49 
are the most luminous Galactic HII regions, we conclude that the M51 sample
spans approximately the full range of Galactic compact HII regions (perhaps
not sampling to the lowest luminosities but that depends on the order of magnitude 
extinction correction adopted for M51). For reference, we also note that the 30Dor cluster in the LMC
is a few times more luminous than W49. The diameter of the 30Dor stellar
cluster is $\sim$ 20 pc and the associated H$\alpha$ emission 
extends over $\sim$ 100 pc.
In the full sample of M51 regions
(Fig.~\ref{lf_comparison_same}), the most luminous is at an {\it observed} L$_{H\alpha}$ = 2$\times 10^{39}$ erg s$^{-1}$; if this region has a modest extinction, it's 
luminosity would be 20 times that of W49. This is another reason for believing 
that the most luminous H$\alpha$ regions in M51 are blended.

Although the luminosity range of M51 HII regions with D $\leq$ 50 pc is similar to that 
of the Galactic regions listed in Table 3, their sizes are typically
several times larger and their mean densities lower. These differences have several likely
explanations : the Galactic regions were selected for high surface
brightness free-free emission (ie. compact radio HII regions) in
order to avoid possible extended, missing flux; the M51 regions are
still only marginally resolved and undoubtedly would include 
compact cores at higher resolution; or some of the Galactic regions
may be in an earlier stage of their evolution before they have fully 
expanded. Despite these differences, the ionizing luminosities are
similar and it is those luminosities which we use in the next section to probe
the OB star cluster mass distributions and temporal evolution in M51.

The luminosity function of Galactic radio HII regions has been fit 
by Smith \& Kennicutt (1989) and McKee \& Williams (1997) who found power law 
indexes of -1.3 and -1.0, respectively -- consistent with our value of -1.01 for M51.  
Although McKee \& Williams (1997) derive an upper limit for the Lyman continuum emission rate from Galactic HII 
regions of Q$_{LyC}$~$_{up}$ = 4.9$\times$10$^{51}$ s$^{-1}$ (including a correction of 25\% for absorption by internal dust), the actual maximum observed (W49) is only $7\times 10^{50}$ s$^{-1}$. The latter value is consistent with our M51 extinction-corrected upper 
limit Q$_{LyC}$~$_{up}$ $\sim$ 7$\times$10$^{50}$ s$^{-1}$ (eq. 10, from the 50 pc sample). Our estimate has no corrections 
for dust absorption of the ionizing photons or UV escaping into the diffuse medium. The latter factor adpted in the McKee \& Williams (1997)
study is very large (1/0.29 = 3.45). They obtained this number by assuming the 
entire discrepancy between the far-infrared COBE measurements of [NII]
and the sum of the Ly continuum from discrete HII regions was due to 
leakage of ionizing UV out of the regions measured in radio free-free. 
Alternatively, much of the discrepancy might be due to older or lower mass OB associations 
(which are undetected in the Galactic radio sample), ionization by other sources or 
even uncertainties in the analysis of the COBE data. We do not include this factor in view of its large uncertainty and because we wish to compare {\it  discrete} HII regions in M51 with comparable objects in the Galaxy.

\section{OB Star Clusters and HII Region Evolution}

In order to understand the nature of the constraints provided
by the observed HII region H$\alpha$ luminosities and their distribution, 
we develop here a simplified model for OB star clusters 
and HII region evolution. The critical constraints 
which we wish to understand are : 

1) the slope of the luminosity function on the high luminosity end (d~N(L)/d~ln~L $\propto $L$_{H\alpha}^{-1.01}$ , Eq. 1) ---  to what extent is this due to the 
evolutionary decay of the Lyman continuum emission from clusters as they age,
as opposed to the mass spectra of the clusters and their high-mass
stars; and   

2) the existence of an 'approximate' maximum luminosity for individual
OB star clusters at L$_{H\alpha}$ $\sim 10^{39}$ erg s$^{-1}$. (This maximum
is surprising in view of the fact that it corresponds to a cluster
of only a few $\times 10^3$ \msun~ yet GMCs contain 10$^{5-6}$ \msun~ of molecular gas;
thus, there is  sufficient mass available to form much more massive clusters.)

\subsection{Model Stellar and Cluster Parameters}

For each OB
star cluster, we assume the star formation occurs on a timescale short compared to stellar evolution timescales.  We adopt a Salpeter initial
mass function (IMF) with N(m$_*$)dm$_*$  $\propto m_*^{-2.35}$ over the range m$_l$  ---  m$_u$ (here, taken to be 1 and 120 \msun, respectively) rather than the local Galactic IMF (\cite{sca87}). The Salpeter IMF is consistent 
with the recent determinations for OB associations which have power-law 
indexes in the range $-$2.1 $\to$ $-$2.45 (\cite{mas95}) and the 30Dor 
cluster in the LMC (\cite{sel99}). 

The stellar lifetimes and Lyman continuum emission 
rates are based on stellar models with constraints provided
by observations of individual high-mass stars. 
Main sequence lifetimes and bolometric luminosities are from Renzini and Buzzoni (1986)
and Maeder (1987) for $\leq 10$\msun~ and $\geq 10$\msun,~respectively.  For reference, the MS lifetimes are 8.6, 4.5, 3.8 
and 3.3 $\times10^6$ yr for 20, 40, 60 and 80 \msun. The ionizing photon production rates (Q$_{LyC}$) are from 
Vacca, Garmany \& Shull (1996). These are shown in Fig.~\ref{star}. 

In building up the stellar population of a star cluster, we take advantage 
of the fact that the stellar IMF should be interpreted as a probabilty
distribution and that even the low mass clusters will have the same proportion of high and low mass stars, independent of the cluster mass (subject to the condition that a very low mass cluster can't have a star of mass greater than that of the cluster). Thus, if one samples enough low mass clusters, their average population is the same as a single high mass cluster with the entire range of stellar masses sampled. An ensemble of lower mass clusters will
therefore have the same Lyman continuum ouput per unit cluster mass as a higher mass 
cluster with a 'saturated' IMF (Oey \& Clarke 1999). For this reason, if 
one is interested in the behavior of the high luminosity tail of the luminosity function (and not the dispersion of the luminosity function at lower 'unsaturated' luminosities), one can calculate the evolution of a single 
'saturated' cluster and scale the properties directly with cluster mass
to derive the mean properties of lower mass clusters.  

\subsection{Evolution of Cluster Luminosity and Q$_{LyC}$}

Once assembled, we track the 
evolution of the cluster Lyman continuum and luminosity, removing high mass
stars after their MS lifetimes. 
In Fig.~\ref{q_l_age} the 
temporal evolution of the Lyman continuum emission rate and total luminosity 
are shown scaled to a cluster mass of 10$^3$ \msun (between 1 and 120 \msun).
For lower or higher mass clusters the mean Q$_{LyC}$ and L can be scaled 
linearly with mass (as long as one is analyzing a large population of clusters). 
Assuming formation of all stars in the cluster in a time short compared to the stellar evolution times, the Lyman 
continuum emission rate will remain constant until an age equal to the 
main sequence lifetime of the most massive stars ($\sim 3 \times 10^6$ yrs -- see above and Fig.~\ref{star}). Subsequently, Q will decay exponentially in time 
with an e-folding time of $\sim 10^6$ yrs out to 10$^7$ yrs at which point the 
Ly continuum is relatively insignificant (Fig.~\ref{q_l_age}). 

Integrating over the lifetime of the 10$^3$ \msun ~cluster, the total number of Lyman 
continuum photons is 1.2$\times 10^{64}$. This implies 1.2$\times 10^{61}$ Lyman continuum photons per solar mass of stars between 1 and 120 \msun ~ or 
equivalently 4.8$\times 10^{60}$ Lyman continuum photons per solar mass of stars between 0.1 and 120 \msun. The latter estimate is 65\% greater than that dervied by Kennicutt, Tamblyn \& Congden (1994) for the same IMF and mass range. The agreement is quite
acceptable given the fact that their relation was dervied using Kurucz (1992)
model atmospheres rather than the Vacca \etal 1996 observationally-constrained 
ionizing outputs which are generally higher than Kurucz's.

Thus for a steady state star formation rate SFR, the OB clusters
yield 

$$ Q_{LyC} / SFR = 1.52\times 10^{53} sec^{-1} (M_{\odot} yr^{-1})^{-1}   \eqno (11) $$

\noindent for a Salpeter IMF between 1 and 120 \msun.  This result is useful for 
estimates of time- and global-averaged star formation rates (see \Section{14}).

\subsection{Cluster Masses}

The extinction-corrected H$\alpha$ luminosities for the HII region sample with diameters $\leq 50$ pc (\Section{11}) imply maximum Lyman continuum emission rate of Q$_{LyC}$ = $7\times 10^{50}$ s$^{-1}$. 
This corresponds to cluster mass of 7$\times 10^3$ \msun~(see Fig.~\ref{q_l_age}) 
and a typical mean mass is $\sim 10^{3}$ \msun. 
 Above approximately 10$^3$ \msun, the cluster contains a reasonable sampling of all stellar masses and the stellar population is said to be 'saturated' (cf. Oey \& Clarke 1999). The Lyman continuum production
rate grows linearly for further increases in the cluster mass.  If the Salpeter IMF is extended
down to 0.1 \msun, the total cluster mass estimate is increased by a factor 
of 2.5. 

We should stress that given the large number of HII regions occupying the 
unsaturated (flat) part of the luminosity function, estimates 
of the {\it individual} cluster masses for these unsaturated HII regions is clearly 
very risky based on H$\alpha$ luminosities (or any Lyman continuum measure).
Since their stellar populations are poorly sampled, the ratio of UV
output to total stellar mass must fluctuate greatly from
one unsaturated region to another. And of course, since these regions
are selected for having detectable H$\alpha$, they will be biased towards high ratios of LyC per solar mass of stars.  

\subsection{HII Region Luminosity Functions}

In order to model the observed HII region luminosity function, one
must sample the modelled OB star clusters over a range of cluster masses and over the lifetime of the OB stars. Assuming approximately coeval star formation, the temporal evolution of the Lyman continuum luminosity from the clusters is determined by stellar evolution of the 
highest-mass stars. The luminosity function of the HII regions can then 
be modelled by sampling uniformly in time the Lyman output from 
each cluster mass, weighted by the initial mass distribution of clusters.

In analogy with the treatment of stellar IMF,
we assume that the initial {\bf cluster} masses can also be represented 
as a power-law :

$$ {N(M_{cl}) d~M_{cl}}  \propto M_{cl}^\Gamma . \eqno (12) $$

\noindent In Fig.~\ref{lf_th} the luminosity 
functions are plotted for 5 values of the mass power law index $\Gamma$ between -1 and -3 (Eq. 12). The slope of high luminosity end of
the luminosity function is directly determined by the cluster mass spectrum index $\Gamma$ (McKee \& Williams 1998) since the highest luminosities are 
contributed by 'saturated' clusters during their initial constant luminosity
phase ($< 3\times 10^6$ yrs). The model luminosity functions shown in Fig.~\ref{lf_th}
clearly show that flat or nearly flat distribtions of cluster mass
are ruled out by the observed luminosty functions. An excellent match to the observed luminosity function power law index -1.01 is thus provided 
by N(M$_{cl}$)dM$_{cl}$ $\propto$ M$_{cl}$$^{-2.01}$ with variations of $\Delta\Gamma = \pm0.15$ between
arm and interarm regions. The derived power law index for the cluster mass spectrum is close to but not identical to that ($\Gamma \simeq -1.6$, \cite{sco78}) of 
Galactic GMCs.

Thilker \etal (2001) did a 
similar analysis of their H$\alpha$ measurements to obtain an initial cluster mass spectrum N(M$_{cl}$)dM$_{cl}$ $\propto$ M$_{cl}$$^{-2}$
(after exploring a range of $-1.75 \rightarrow 2.25$). Both ours and their results provide strong evidence of a steeply falling cluster mass spectrum. 
These cluster mass spectra are also in excellent agreement with the value $\Gamma$ = -2 derived by 
Larsen (2000) and Bik \etal (2001) from UBV photometry of the clusters in M51 (as opposed to the ionized gas). 

The luminosity functions shown in Fig.~\ref{lf_th} were computed for 
a cluster mass range between 500 and 5000 \msun. The adopted upper limit in the 
cluster mass produces the cutoff in the luminosity function at Q$_{LyC}$ = 4$\times 10^{50}$
s$^{-1}$. For cluster with a 'saturated' IMF, the location of this 
break will scale linearly with the upper limit to M$_{cl}$. The upper limit to cluster mass is 
$\leq 7000$ \msun (between 1 and 120 \msun) for the D $\leq$ 50 pc sample, assuming a factor of 10 increase for extinction. Moreover, since some of the most 
luminous regions in the sample are possibly blends, the real upper mass limit
is probably less. 

\subsection{Upper 'Cutoff' of Cluster Mass}

Here we briefly explore the possibility that there is a physical mechanism limiting the buildup of clusters more massive than a few $\times 10^3$ \msun.
In the following, we track the 
feedback of radiation pressure on buildup of a stellar cluster in a molecular cloud core. The important role of radiation pressure in high-mass star formation has not been amply recognized. 

We posit a simple model for cluster buildup in which the protocluster
forms at the center of a molecular cloud core in which the density 
rises with decreasing radius.  We assume that initially the entire cloud 
core is in free-fall collapse.  The stars within the growing cluster have a Salpeter IMF  ---  thus since lower mass stars form with a higher probability the initial low mass cluster will have few high mass stars. However, as the cluster grows and becomes more massive,
the IMF is populated to higher mass. The 
surrounding gas feels the gravitational 
attraction of the interior mass M$_R$ (stars plus the interior 
core gas). Lastly, we assume that  
gas accreted inside the adopted cluster radius is added to the cluster and the total cluster mass is instantaneously redistributed with the Salpeter IMF. 

This approach thus neglects the complication that low or high mass stars might 
preferentially form at different epochs due differences in the
physics of their formation or their different pre-main-sequence evolutionary times. However, we should note that 
the assumptions made by us do not really require that all stars form at the same
time but rather that the prestellar condensations which become stars of different mass
have formed and accreted into the cloud core region. Since the low-mass stars
have low luminosity, they are only important for their gravitational mass
as far as our model is concerned. 
Thompson \etal (1998) find evidence that the low and high mass star formation 
in NGC 2264 is coeval to within 10$^3$--10$^4$ yr. 

Initially, the cluster will contain only a few low-mass
stars and the gas dynamics will be entirely determined by the self-gravity 
of the cloud core and cluster. Since we neglect rotational or magnetic 
stresses, this is clearly the {\bf most favorable} situation for 
accretion and maximal growth of the cluster. On the other hand, as the cluster 
becomes more massive and 
populates the upper main sequence, the higher luminosity-to-mass ratio 
of high-mass stars will result in increased radiation pressure 
on the surrounding dust --- eventually terminating further accretion to the cloud core. The outward radiation pressure will dominate self-gravity at radius R when 

$$ {L< \kappa >\over 4\pi R^2 c } \ge {G M_R \over R^2 } 
 ~~~ , \eqno (13) $$

\noindent where $< \kappa >$ is the {\bf effective} radiative absorption coefficient 
per unit mass. 

Although the original stellar radiation 
is primarily UV and visible, the dust in the cloud core absorbs 
these photons and reradiates the luminosity in the infrared. The 'effective'
absorption coefficient takes account of the fact that outside the radius
where A$_V$ $\sim 1$ mag, the luminosity is at longer wavelengths where the 
dust has a reduced absorption efficiency.  For the standard ISM dust-to-gas ratio (\cite{boh75}),
A$_V$ = 1 mag corresponds to a column N$_H$ = 2$\times10^{21}$ cm$^{-2}$.
We therefore adopt  

$$ < \kappa > = 312 ~{\lambda _V \over \lambda_{eff}(R)} ~cm^2 gr^{-1}
 ~~~  \eqno (14) $$

\noindent and $\lambda_{eff}(R)$ is the absorption coefficient-weighted mean wavelength
of the radiation field at radius R and we adopt a $\lambda^{-1}$ variation of the absorption efficiency with wavelength.

Combining Eq. 13 and 14, we find that the radiation pressure will exceed the gravity of the 
cluster stars when 

$$ ( L / M )_{cl}~ \ge 42  {\lambda _{eff} \over \lambda_{V}} {\lsun \over \msun}
 ~~~ . \eqno (15) $$

\noindent If $\lambda_{eff}$ $\sim$ 3 $\mu$m, $\lambda_{eff} / \lambda_{V} \sim$
10. Thus, for clusters with luminosity-to-mass ratios exceeding $\sim$ 500 \lsun
/ \msun, radiation pressure will halt further accretion. This luminosity-to-mass ratio is reached at about the point when the upper main sequence is first  
fully populated, i.e. a cluster with 
approximately 2000 \msun~ distributed between 1 and 120 \msun. In the above discussion, we conservatively adopt luminosities 
corresponding to the main sequence rather than the much larger, short term 
pre-MS luminosities. This assumption is thus most conservative in 
estimation of the radiation pressure effects. We have also assumed the 
density of dust and gas to be only a function of radius. If the material is 
instead contained in optically thick clumps (possible individual pre-stellar
condensations) or a disk, then the strength of the radiation pressure compared to gravity 
is reduced by a factor which depends on the opacity of the clumps and their 
areal covering factor at each radius. Since neither of these factors 
are constained by existing observations, we have adopted the simplest case
of uniform areal coverage and spherical symmetry. In reality, the gas is likely to 
be clumped and there will be multiple core regions within each GMC. The latter would 
certainly be unresolved in ground based H$\alpha$ imaging and probably even at the 
higher resolution used here. On the other hand, it may be that even if there are multiple cores,
not all will have their cluster formation synchronized to within 10$^5$ yrs (the timescale 
of the phenomena we have been discussing). 

In the above, we assumed the most favorable conditions for 
cluster growth --- free-fall collapse and no rotational or magnetic 
impediments. Additional
outward pressure can of course arise from the hot ionized gas associated with the OB stars and their stellar winds; these effects are undoubtedly
important later in the evolution of the cluster but they are also more dependent on the 
HII region geometry. By contrast, radiation pressure is inescapable 
during the early phases of cluster formation and therefore must
 play a central role in regulating the initial cluster growth.  The subsequent
dynamic evolution of cloud core will be affected by 
all three : radiation pressure, HII expansion and stellar winds.  
  
Elmegreen (1983) has also pointed out the importance of radiation pressure from a forming
star cluster in possibly limiting the maximum mass of star clusters. He used somewhat 
different numerical values for the dust opacity and did not consider the reduced 
effective dust absorption cross section due to the reprocessing into the infrared. 
Although he used a higher effective dust cross section, this was compensated by
his assumption that there will be substantial gas mass in the intracluster space and he 
arrives at a similar limiting cluster mass of $\sim 10^3$ \msun.

\subsection{OB Star Cluster Formation and Expanding Associations}

To understand better the limit on the core stellar population of OB star clusters, we have numerically modelled the collapse of a cloud core
with a growing star cluster. The cloud core, having an initial power-law radial density profile,
 
$$ n =  n_0 \left({ R \over R_{cl} }\right)^{-2}
 ~~~ , \eqno (16) $$

\noindent is assumed to start in free-fall collapse with
V$_R$ = V$_{ff}$ = -(2GM$_R$/R)$^{1/2}$. In the subsequent dynamic evolution of the cloud core,
envelope gas getting inside the adopted cluster radius R$_{cl}$ (taken to 
be 0.01 pc) is added to the cluster with an efficiency of 50\%. At each time step, the 
cluster mass
is distributed into stars with a Salpeter IMF and the cluster luminosity updated. 

To compute the radiation pressure at each radius
we adopt an effective wavelength for the radiation from 

$$ \lambda _{eff}(R) \simeq { T_V  \lambda _{V} \over (1-e^{-\tau _V(R)})T_D(R)+T_{cl} e^{-\tau _V(R)} }
 ~~~ , \eqno (17) $$

where the dust temperature is given by

$$ T_D(R) = 70 \left({L \over 10^5 \lsun}\right)^{1/5} \left({2 \times 10^{17} \over R }\right)^{2/5} K  
 ~~~ . \eqno (18) $$

\noindent This scaling with L and R is based on infrared observations of the 
OMC-1 cloud core (Werner \etal 1976). Eqs.~15 and~16  provide a reasonable approximation to the variation of dust temperature 
in an {\bf optically thick} cloud heated by a central luminosity source. Due to the high opacity of the cloud core, the dust is heat mainly by reradiated  emission from interior shells (as opposed to direct stellar photons). 
For the effective
temperature of the cluster luminosity, we adopt a constant value T$_{cl}$ = 30,000 K, independent of 
the cluster mass. 

The Lagrangian form of the hydrodynamic equation of motion was advanced in time with gravity, thermal and radiation pressure terms. Mass shells were distributed logarithmically with the less massive shells on the inside in order that the transition from $\tau _V$ = 0 $\to$ 5 was well resolved. 
Artificial viscosity with a coefficient of 3 was introduced to stabilize the radiatively compressed shells (\cite{chr67}). The gas was taken to be isothermal with an
'effective' sound speed of 1 \kms. The initial density scale 
had n$_0$ = 10$^8$ cm$^{-3}$ at 0.01 pc and the cloud core extended out to a radius such that its total mass was 2$\times10^4$ \msun. 

The density and velocity profiles at 50,000 yr intervals 
after the initial free-fall collapse are shown in Fig.~\ref{clu}.
As expected from the discussion above, the radiation pressure becomes 
dominant over gravity for the inner shells of the infalling
gas when the cluster has reached $\sim$ 500 \msun. At this point, 
the infall of the inner shells is reversed and they rapidly 
accelerate outwards and collide with the still infalling, exterior shells.
The outer shells have high dust opacity (to the central cluster) 
and therefore are not subjected to such high radiation pressures. 

Where the outward and inward moving gas collide, a dense, 
shock-compressed shell 
forms and accelerates outwards.  Typical outward radial velocities
are $\sim$ 2 to 6 \kms ~and the density of the shocked layer is 
greatly enhanced over that of the initial density profile (Fig.~\ref{clu}). At this point, further accretion to the cluster core 
is effectively shut off. For the parameters used in this calculation, 
the cluster ended up with 881 \msun~ after 500,000 yr. Without radiation 
pressure, free-fall collapse with the same parameters would have produced a cluster mass of 5800 \msun~
in the same interval.

It is interesting to note that the outward moving shell may propagate 
a second phase of stimulated star formation into the cloud envelope.
The compressed gas in this shell can be unstable, collapse and fragment into a second generation of stars. The density of the compressed layer is typically enhanced a factor of 100-1000, becoming 10$^{5-6}$ cm$^{-3}$ 
at radii of a few pc. At T = 50 K these densities yield a Jeans length of $\sim$ 1/3 pc and mass of 50 \msun. This second-generation star formation could become 
a significant enhancement to the initial cluster core mass provided 
the density scale n$_0$ is high enough. Eventually the density in the outward moving shell will drop due to spherical divergence and the falloff of the outer envelope density. The shell would then no longer be unstable and stimulated star formation would stop. Stars formed within 
the expanding shell will of course have a net outward radial velocity 
of a few \kms and be unbound from the core cluster.    
Evidence of radial expansion in OB associations is seen in 
the I Orion subgroup d (Trapezium cluster; \cite{bla64}) with
a radial expansion of 2.5 \kms ~at the outer edge (R$\sim$ 0.7 pc) of the cluster. It would be interesting to analyze the Hipparcos data for nearby OB 
star clusters to look for evidence of similar expansions. Dreher \etal (1984)
discovered a ring (R $\sim$ 0.4 pc) of ultra-compact HII regions in the W49 complex, each of which is ionized by an internal star or star cluster;  
 this ring of high mass star formation 
might correspond to the second wave of triggered star formation 
discussed above. 

Recent observations of 
luminous IR galaxies have shown considerably more luminous (presumably more massive) super star clusters (SSC; eg. \cite{whi99}). In the context of the 
discussion above, it might be speculated that such clusters could form
as a result of a particularly prodigious second-wave of triggered star formation in unusually
massive and dense molecular clouds. The galaxies hosting SSCs are largely interacting starburst systems (eg. M82 and the 'Antennae' galaxies)
in which the molecular clouds are likely to be both more massive and 
of higher density. Clearly, if the density remains high further out 
from the star cluster core, the triggered wave of star formation can 
propagate further and generate a more massive and more extended cluster.     
 
A similar scenario for stimulated
star formation in successive OB associations was suggested by Elmegreen \& Lada
(1970). In their model, the compression wave was due to expansion of the 
hot, ionized Str$\ddot{o}$mgren sphere rather than radiation pressure and the
motivation was to account for the temporal sequence of {\bf separate}
OB associations. Their model is different in both the 
physical mechanism for gas compression (HII region shock fronts) and in geometry --- separate OB associations formed in the compressed shell surrounding 
a large, expanding HII region.  The model described above is discussed in more detail
in a forthcoming paper.

A different scenario for the formation of SSCs has been modeled by Tan \& McKee (2000).
They suggest that in a clumpy GMC with multiple cloud core regions each of a few $\times 10^3$ \msun, the SSC might form simply from have a simultaneously high efficiency of star formation in 
all of the cores and the SSC is then the composit of the multiple star clusters and whether 
it is gravitaionally bound or not then just depends on the overall efficiency of star formation
for the entire GMC and consequently an assumption that a large fraction of the 
original GMC mass is contained in the core regions as opposed to the intraclump medium.
(It is worth noting, that the individual cores or clumps in their model are posited
to have mass of a few $\times 10^3$ \msun\ so the clusters forming in each 
core still obey the radiation pressure limit discussed above.)

\section {Total LyC and Star Formation Rates}

The total H$\alpha$ luminosity can be used to estimate the overall Lyman continuum output 
and formation rate of OB stars in M51. In doing this, we separate 
the diffuse and discrete HII regions and we must make corrections for the outer galaxy not covered in the WFPC2 images. In Table 4,
we summarize the observed H$\alpha$ luminosities, the adopted 'typical' extinctions, extinction-corrected H$\alpha$ luminosities and implied Lyman continuum outputs for the
separate components and regions of the galaxy. 

For the discrete HII regions, we adopt 
A$_V$ = 3.1 mag (see \Section{10}) and therefore increase their luminosity by 
a factor of 10 (Eq. 8). For the diffuse gas, the extinction is probably less and we adopt A$_V \simeq$ 1 mag and thus increase the diffuse luminosity by a factor of 2.1 (Eq. 7). (The extinction of the diffuse gas could not be 
reliably estimated from our P$\alpha$ since it would require extremely 
accurate determination of the backgrounds. The adopted 
1 mag is simply based on the presumption that since the diffuse gas is more
widespread, both between the dust clouds and at higher scale height, its extinction is lower than for the discrete HII regions measured here.) The H$\alpha$ emission in the outer galaxy which was not covered in the WFPC2 images was estimated by multiplying the luminosity measured in the 
WFPC2 area by a simple  scale factor (= 1.92). This factor is equal to the ratio of the total observed H$\alpha$ flux in M51 (\cite{ran92}) to that measured by us in the WFPC2 area.

The total Lyman emission rates for the WFPC2 area and M51 are 3.0 and 7.3$\times 10^{53}$ s$^{-1}$. The steady-state star formation rate 
as 2.00 and 4.79 \msun ~yr$^{-1}$ of stars between 1 and 120 \msun for the WFPC2 and total M51 areas (using Eq. 11). 
The inferred rates are increased by a factor of 2.5 if the IMF is extended down to 0.1 \msun. 

The total mass of star forming gas can be estimated from single dish
CO observations (eg. \cite{sco83}). The H$_2$ mass within 4 kpc radius 
is approximately 3$\times 10^9$ \msun~and the total for the galaxy is 
6$\times 10^9$ \msun~(\cite{sco83} after scaling for a conversion factor of  N$_{H_2}$/I$_{CO}$ = 2.2$\times 10^{22}$ cm$^{-2}$ / K \kms ~rather
than 50\% larger value used there). The implied cycling times for a typical H nucleus 
to pass into a new generation of stars is therefore 1.2$\times 10^9$ yr
for both the inner galaxy and total M51. On the one hand, this time would be shortened if the star formation associated with the HII regions 
extends below 1\msun; on the other hand, 
approximately 50\% of the mass absorbed in the stars actually gets recycled 
back into the ISM through stellar evolution on a timescale of 10$^9$ yr (see \cite{nor88}). 

\section {Conclusions}

High resolution HST imaging of H$\alpha$ and P$\alpha$ has been used to 
analyze the properties of HII regions and OB star formation in M51. The critical 
aspects of these data are : the high resolution (4 --- 9 pc as compared with 
$\sim$100 pc in earlier ground-based imaging) which enables the clear 
separation of individual star formation regions, the ability to 
determine and correct for dust extinction using the H$\alpha$/P$\alpha$ 
flux ratio and the high sensitivity which enables us to probe 
HII regions with luminosity well below that of M42 (the Orion Nebula).
From the observational data we find :

1) A total of 1373 H$\alpha$ emission regions were defined (using 
automated procedures). The total flux of these discrete regions constitutes
about 31\% of the total H$\alpha$ emission in the central 281 $\times$ 223 \arcsec\ 
of M51. The observed H$\alpha$ luminosities range from 10$^{36}$ to 
$2\times 10^{39}$ erg s$^{-1}$ and their diameters are mostly from 10 to 100 pc
with a mean value of $\sim$30 pc. (For the higher luminosity and usually larger regions, 
the mean electron density is generally lower, strongly suggesting that some of these
regions are likely to be blends of multiple HII regions.) 

2) The observed H$\alpha$ luminosity function exhibits a broad peak 
at L$_{H\alpha}$ $\sim 10^{37}$ erg s$^{-1}$, falling 
as L$_{H\alpha}$$^{-1.01}$ on the high luminosity side. No
evidence is seen for the break in the luminosity function at 
Log(L$_{H\alpha}$)=10$^{38.6}$ erg s$^{-1}$ reported in ground-based studies. Compared with ground-based determinations, the luminosity function derived here 
is much less populated at high luminosity  ---  very likely most of the regions above 10$^{39}$ erg s$^{-1}$ seen in ground-based
studies were blends of multiple lower luminosity regions which are 
separated here.

3) Significant differences are seen in the luminosity functions 
of spiral arm, interarm and nuclear HII regions with the spiral arm luminosity functions being flatter.

4) The observed correlation between HII region luminosity and apparent 
size (L $\propto D^{2}$) strong implies that the high luminosity 
regions are blends or superpositions of multiple lower luminosity regions. 

5) For 209 regions which had $\ge$ 5$\sigma$ detections in both P$\alpha$
and H$\alpha$, the observed line ratios were used to derive the 
overlying visual extinction assuming standard Galactic dust extinction curves and intrinsic line ratios given by Case B recombination. The implied 
extinctions range from A$_V$ = 0 to 6 mag, with an intensity-weighted mean value of $<A_V>$ = 3.1 mag. Thus the observed H$\alpha$ luminosities 
should be increased by an average factor of 10, implying similar increases in 
the OB star cluster luminosities and implied star formation rates. The high extinctions
 underscore the need for extinction estimates in analyzing the 
properties of HII regions and their associated OB star clusters. 

6) The extinctions are also highly variable from region
to region and across individual regions. (In deriving the estimates quoted above,
the line ratios were computed pixel by pixel and then averaged over each region.)

7) The most luminous regions have sizes $\ge 100$ pc and they are undoubtedly blends of multiple regions. This is clear since their sizes
are much larger than the maximum diameter ($\leq 50$ pc) to which an HII region might conceivably 
expand within the $\sim 3\times 10^6$ yr lifetime of the OB stars and it is 
also consistent with observed correlation (L $\propto D^{2}$) found between the 
measured luminosities and sizes of the HII regions.

8) A subsample of 1101 regions with sizes $\leq 50$ pc therefore 
constitutes those regions which might conceivably be ionized by a single cluster.
Their extinction-corrected luminosities range between $2\times 10^{37}$ and $10^{39}$ erg s$^{-1}$ (with no corrections for dust absorption of the Lyman continuum 
or UV which escapes to the diffuse medium). The range is roughly comparable to 
HII regions ranging between 2/3 of M42 (the Orion Nebula) and W49 (the most luminous Galactic radio HII region). The upper limit for 
individual cluster HII regions is therefore conservatively $\leq 10^{39}$ erg s$^{-1}$.

In additon to the observational data, we have modelled the Lyman continuum 
luminosity as a function of total cluster mass with stars distributed 
with a Salpeter IMF (1  ---  120 \msun) and as a function of cluster age. 
This model is used to derive the cluster properties. We find :

1) The upper limit to the $\leq 50$ pc sample luminosity function 
corresponds to an ionizing photon production rate Q$_{LyC}$~$_{up} \simeq 7\times10^{50}$
s$^{-1}$ and the corresponding upper limit for cluster masses is $\leq $5000 \msun. (The implied masses are increased by a factor of 
2.5 if the Salpeter IMF is extended down to 0.1 \msun.)

2) For a power-law distribution of cluster masses, 
we find that the observed peak in the luminosity function at 10$^{37}$
erg s$^{-1}$ clearly rules out flat or slowly falling 
cluster mass spectra. The observed $-$1.01 power-law index on the 
high luminosity tail of the luminosity distribution implies a 
cluster mass spectra N(M$_{cl}$)/d~M$_{cl} \propto M_{cl}^{-2.01}$.

3) The highest mass clusters are approximately $\sim $1000 \msun. The parent molecular clouds are much more massive and one must ask : why don't the OB star clusters generally build up to much greater mass? We suggest that the formation of a massive cluster in a molecular  
cloud core is likely to be terminated at the point when the luminosity-to-mass ratio 
of the cluster reaches $\sim$1000 \lsun/\msun ~when radiation pressure
begin to dominate the self-gravity of the cluster. This is roughly the point at which the 
Salpeter IMF first becomes fully populated.

4) A hydrodynamic model with an initial R$^{-2}$ density distribution in free-fall collapse verifies that the core star cluster is likely to self-limit at $\sim$10$^3$ \msun. At this point, radiation pressure effectively 
terminates further gas and dust accretion to the central core. However, we also find 
that a radiatively-compressed shell will then propagate outwards at a few \kms, possibly triggering a second wave of star formation out to a few pc radius. This may, in fact, be the mechanism for forming the more massive star clusters (SSC) seen in starburst galaxies 
where the densities are likely to be higher out to larger radii in the cloud envelopes. The stars
formed in the expanding shell will have significant outward radial 
motion and will probably be unbound.

Lastly, we have combined our modelling results with measurements 
of the total line emission to evaluate the total Lyman continuum output and global star formation 
rate. From the extinction-corrected H$\alpha$ luminosity, we find 
 Q$_{LyC}$~$_{total}$ = 7$\times10^{53}$ s$^{-1}$ and SFR(1-120\msun) = 4.17 \msun~yr$^{-1}$.

\vspace{5mm}

The
NICMOS project has been supported by NASA grant NAG 5-3042 to the NICMOS
instrument definition team.  This paper is based on observations with the
NASA/ESA Hubble Space Telescope obtained at the Space Telescope Science
Institute, which is operated by Association of Universities for Research
in Astronomy, Incorporated, under NASA contract NAS5-26555. We are very grateful to Rich Rand for several discussions during the course of our work 
and for permitting us access to his H$\alpha$ images. We also thank
David Thilker for sharing his HII region photometry software with us and for
extremely constructive and thorough refereeing of this paper. We also thank Jonathan Tan, Chris McKee and Crystal Martin for very useful comments and 
suggestions on the paper. We thank Zara Turgel for careful proofreading and comments on the manuscript.

\clearpage

\figcaption[halpha_large.ps]{The full mosaic of WFPC2 H$\alpha$ (continuum subtracted) images for the central 281 $\times$ 223 \arcsec\ of M51. The H$\alpha$ emission is shown in red and continuum V and B bands are in green and blue, respectively. (In the red, we also added the I band image in order to balance the colors on the stars.)
\label{halpha_large}}

\figcaption[spiral.ps]{The flow streamlines are shown for material at 3, 3.2, 6 and 6.2 kpc 
in a spiral potential like that of M51 based on the model of Roberts \&
Stewart (1987) with 
 the spiral arms starting at R = 2 kpc and PA=15\deg.  Since the 
pattern speed is approximately half the circular velocity, the gas approaches the spiral arm
from the underside, is deflected along the arm toward smaller radius and then leaves the arm
on the front. Orbit crowding is due to the  
spiral arm streaming motions. The cross-marks correspond to 250 Myr time intervals. 
\label{spiral}}

\figcaption[palpha_large.ps]{The mosaic of NIC3 P$\alpha$ (continuum subtracted) images covering the central 186 $\times$ 188 \arcsec\ of M51. The P$\alpha$ emission is shown in red and continuum V and B bands are in green and blue, respectively. (In the red, we also added the I band image in order to balance the colors on the stars.)
\label{palpha_large}}

\figcaption[halpha_palpha.ps]{The H$\alpha$ (red) and P$\alpha$ (green) images are shown for the 
area of overlap. The blue was made by the combination (V-H)+V in order to show the dust lanes and the background disk stars. Areas with yellow color 
have strong H$\alpha$ and P$\alpha$.
\label{halpha_palpha}}

\figcaption[halpha_palpha_small.ps]{The H$\alpha$ (red) and P$\alpha$ (green)
are shown for the area to the west of the nucleus with all continuum removed.
Although virtually all reasonably bright H$\alpha$ emission regions 
are seen in P$\alpha$, extremely large variation in the line ratio
are apparent from region to region and the morphology of individual regions
can be very different in H$\alpha$ and P$\alpha$.
\label{halpha_palpha_small}}

\figcaption[halpha_comparison.ps]{The H$\alpha$ emission structure apparent 
at 0.1\arcsec using HST (left) is compared with the same 
image smoothed with a 1.5\arcsec\ FWHM Gaussian to simulate typical ground-based resolution. The sub-image area shown here is approximately 60\arcsec\
north of the nucleus. The length bar 
corresponds to 100 pc. Galactic GMCs have diameters typically 2 to 5 times smaller and their associated HII regions have still smaller sizes.
This comparison illustrates well that the gound-based and HST imaging
are clearly studying very different physical structures  ---  the former, 
{\bf associations} of OB star formation, the latter, perhaps individual 
OB star formation regions. 
\label{halpha_comparison}}

\figcaption[sub_area.ps]{For the same area shown in Fig.~\ref{halpha_comparison} (60\arcsec\
north of the nucleus), we show contours for H$\alpha$ and P$\alpha$ emissions (top panels). In the lower right panel, the H$\alpha$ (red) and P$\alpha$ (blue) are shown together. It is  clear that significant differences are 
seen in the H$\alpha$ and P$\alpha$ due to the 
higher extinction in H$\alpha$.  The areas of the H$\alpha$ emission regions as defined by our automated 
algorithm  are shown (the peak positions are indicated by the black pixels). 
The length bar 
corresponds to 1\arcsec\ or 46 pc. 
\label{sub_area}}

\figcaption[hii_sp_dist_r92.ps]{The locations of the H$\alpha$ emission 
regions selected by our automated procedure are shown together with
those measured interactively by Rand (1992). In general, there is 
very close correspondence between the two samples, although in many instances we have catalogued 
multiple regions where Rand had just one. The outer areas were not covered in 
our images and therefore a number of Rand's HII regions don't have a counterparts in our sample; the converse is true in the nuclear region
which was excluded from Rand's analysis due to blending. There are 17 regions in Rands
sample which do not appear in ours although they are in the area analyzed by both studies. All of
these regions are have peak intensities just below our 6 $\sigma$ criterion for the peak intensity. 
\label{hii_sp_dist_r92}}

\figcaption[two_pt.ps]{The two point angular correlation function 
is shown for centroid positions of the 1373 discrete HII regions (top panel) and pixels (bottom panel) with H$\alpha$ emission exceeding 3$\sigma$ in the 
same  H$\alpha$ image with diffuse background subtracted.  The angular sampling of the images was normalized out by dividing the 
correlation functions with those computed for pixels sampled randomly
within the observed image areas. This normalization does not
remove large-scale correlations such as the concentration of HII 
regions toward small galactic radii and along spiral arms; thus neither correlation function 
goes asymptotically to 0 at large angles and their integrals are not equal to zero. 
\label{two_pt}}

\figcaption[nearest.ps]{The percentage of HII regions with nearest neighbors within the specific angular offset ($\theta$) is shown.\label{nearest}}

\figcaption[lf_comparison_same.ps]{The observed H$\alpha$ luminosity function for all HII regions 
defined and measured in the WFPC2 images (not corrected for extinction) is compared with the luminosity functions 
derived from ground-based imaging by Rand (1992) and Petit \etal (1996). The fact that the HST luminosity function appears to shift to lower luminosity 
is due largely to the ability to separate individual HII regions which become blended in ground-based images (see text). Fits to logarithmically binned distributions are over the range  L$_{H\alpha} = 40 \to 800\times10^{36}$ erg s$^{-1}$. Typical extinctions of A$_{H\alpha}$
= 0.798 A$_V$ = 0.798 $\times$ 3.2 will shift the luminosity functions 
a factor of 5.9 higher.  
\label{lf_comparison_same}}

\figcaption[sep_lf_4.ps]{The observed H$\alpha$ luminosity functions separated for arm, interarm and nuclear regions. These areas are defined and shown in Polletta \etal (2001). Fits to logarithmically binned distributions are over the range  L$_{H\alpha} = 12 \to 500\times10^{36}$ erg s$^{-1}$. The power-law fits to the high luminosity tail are also shown. A significant flattening of the luminosity function is seen in the nucleus and spiral arms, compared to the interarm regions. 
\label{sep_lf_4}}

\figcaption[histo_sizes.ps]{The distribution of HII region diameters are shown for regions containing at least 2 pixels, separated between
arm, interarm and nuclear regions. No significant difference is seen in the 
diameter distributions between the different regions. The mean diameters are given for 
all regions smaller than 100 pc. Regions with diameter $\geq$ 120 pc were not 
plotted since these are certainly blends of multiple regions. 
\label{histo_sizes}}

\figcaption[lum_size_density.ps]{The apparent H$\alpha$ luminosities (upper panel) and HII region diameters (lower panel) are plotted as functions of mean electron density
with the arm, interarm and nuclear regions plotted in blue, red, and green, respectively.
\label{lum_size_density}}

\figcaption[lum_size.ps]{The observed luminosities of the HII regions are plotted as a function of HII region diameter.   
\label{lum_size}}

\figcaption[hii_ratio_ext.ps]{Upper Panel : The measured flux ratios (H$\alpha$/P$\alpha$) for 
209 HII regions with boundaries defined from H$\alpha$ (shown in Fig.~\ref{sub_area}) , restricted to those pixels detected at $\geq$ 5$\sigma$ in both P$\alpha$ and H$\alpha$.  For Case B recombination in an ionization bounded HII region with 
no extinction, the intrinsic ratio is 8.15 (dashed vertical line). The lower measured H$\alpha$/P$\alpha$ ratios 
reflect the higher extinction at the wavelength of H$\alpha$ (6563\AA)
than that of P$\alpha$ (1.87 $\mu$m). Lower Panel : The distribution of average visual extinctions for the HII regions inferred from the
measured flux ratios assuming an intrinsic ratio of 8.15 and the standard ISM extinction curve  (Rieke \& Lebofsky 1985, Cardelli \etal 1989). Dashed vertical line is the average extinction A$_V$ = 3.2 mag. (The extinctions were determined at individual pixels and then averaged for each region.)   \label{hii_ratio_ext}}

\figcaption[av_lha.ps]{The derived extinctions as a function of {\it observed} H$\alpha$ luminosity for 
209 HII regions with boundaries defined from H$\alpha$ (shown in Fig.~\ref{sub_area}) 
, restricted to those pixels detected at $\geq$ 5$\sigma$ in P$\alpha$ and H$\alpha$. The extinctions were determined for individual pixels within 
each HII region and then averaged. The solid circles are the averages for 
equal logarithmic bins in H$\alpha$ luminosity.
\label{av_lha}}

\figcaption[Pa_vs_Ha_pxl_th5.ps]{Upper Panel : The measured flux ratios (H$\alpha$/P$\alpha$) for 
all individual pixels detected at $\geq$ 5$\sigma$ in both P$\alpha$ and H$\alpha$.  For Case B recombination in an ionization bounded HII region with 
no extinction, the intrinsic ratio is 8.15 (dashed vertical line). Lower Panel : The derived visual extinctions for the same pixels obtained from the
measured flux ratios assuming an intrinsic ratio of 8.15 and the standard ISM extinction curve  (Rieke \& Lebofsky 1985, Cardelli \etal 1989). \label{Pa_vs_Ha_pxl_th5}}

\figcaption[lf_ext_cor.ps]{The H$\alpha$ luminosities as observed for the 209 regions for which the extinction was derived and as corrected for extinction. The 
extinction-corrected luminosity of each region was calculated by correcting each pixel for 
its extinction and then summing the extinction-corrected pixel luminosities.
These
distributions should not be interpreted as luminosity functions since they include
only those pixels in each region detected at $\ge 5\sigma$ in both lines.
\label{lf_ext_cor}}

\figcaption[lf_50.ps]{The observed H$\alpha$ luminosities plotted for the 1011 HII regions with diameter $< 50$ pc. For the reasons discussed in the text, HII regions larger than 50 pc are almost certainly ionized by multiple clusters; the distribution 
shown here therefore shows the maximum luminosity likely to arise from a single OB star cluster.   
\label{lf_50}}

\figcaption[star.ps]{The adopted Lyman continuum production, stellar luminosity  and main sequence lifetimes 
are shown as a function of stellar mass. The Lyman emission rate is from Vacca, Garmany \& Shull (1996) 
and the main sequence lifetimes and luminosities are from Renzini and Buzzoni (1986)
and Maeder (1987) for $\leq 10$\msun~ and $\geq 10$\msun,
respectively.   
\label{star}}

\figcaption[q_l_age.ps]{The Lyman continumm emission rate and luminosity 
are shown for a 'saturated' cluster as a function of time scaled to 
a total cluster mass of 10$^3$ \msun with a Salpeter IMF from 1 to 120 \msun.
Both the luminosity and Q$_{LyC}$ can be scaled directly proportional to cluster mass for lower and higher mass cluster. 
\label{q_l_age}}

\figcaption[lf_th.ps]{The expected luminosity functions are shown for 5 values of the 
cluster mass power law index ($\Gamma$). The normalized luminosity functions
per logarithmic interval in Lyman continumm emission rate were computed assuming an equal probability of observing a given cluster at any time during its active lifetime. 
\label{lf_th}}

\figcaption[clu.ps]{The density profile for the cloud core, cluster
formation is shown at 0(black), 50(green), 100(turqoise), 150(blue), and 500(red) $\times 10^3$ yr. The inital state was free-fall 
collapse in a R$^{-2}$ density distribution normalized to 10$^8$ cm$^{-3}$ at 0.01 pc. The final mass of the core cluster was 881 \msun ~(compared to 5800 \msun~ for free-fall collapse) and the luminosity was
10$^6$ \lsun. The core accretion was fully halted by radiation pressure within 100,000 yr. After that the radiatively compressed shell moves outward
at 2 --- 6 \kms. Since the compressed shell is likely to be Rayleigh-Taylor and Jeans unstable
and star formation may continue in the outward moving 
shell.
\label{clu}}

\end{document}